\documentclass[12pt]{iopart}
\usepackage{latexsym,epsfig,graphicx,bbm,mdframed}
\usepackage{listings} 
\usepackage[svgnames]{xcolor} 

\mdfdefinestyle{mdfexample1}{userdefinedwidth=1.02\linewidth,skipabove=10pt, skipbelow=10pt}

\definecolor{terracotta}{RGB}{190,75,0}
\usepackage{hyperref}
\hypersetup{
    colorlinks=true,
    linkcolor=DarkRed,
    filecolor=magenta,
    urlcolor=terracotta,
    citecolor=DarkRed,
}

\newcommand{\tnode}{\texttt{tntNode}}
\newcommand{\tnet}{\texttt{tntNetwork}}
\newcommand{\tsys}{\texttt{tntSystem}}
\newcommand{\ket}[1]{\left | \, #1 \right \rangle}

\newcommand{\code}[1]{\texttt{#1}}

\lstdefinelanguage{tnt}
{morekeywords={int, double, tntComplexArray, tntIntArray, tntNode, tntNetwork, unsigned, tntComplex},
sensitive=true,
morecomment=[l]{//},
morecomment=[s]{/*}{*/},
morestring=[b]",
}

\newcommand{\eqr}[1]{Eq.~(\ref{#1})}
\newcommand{\fir}[1]{Fig.~\ref{#1}}
\newcommand{\secr}[1]{Sec.~\ref{#1}}

\newcommand{\braket}[2]{\left\langle\, #1\,|\,#2\,\right\rangle}

\newcommand{\av}[1]{\langle #1\rangle}

\begin{document}

\title[TNT library]{The Tensor Network Theory Library}

\author{S~Al-Assam\dag, S~R~Clark\ddag~and D~Jaksch\dag}
\address{\dag Clarendon Laboratory, University of Oxford, Parks Road, Oxford OX1 3PU, U.K.}
\address{\ddag Department of Physics, University of Bath, Claverton Down, Bath BA2 7AY, U.K.}
\ead{s.r.clark@bath.ac.uk}

\begin{abstract}
In this technical paper we introduce the Tensor Network Theory (TNT) library -- an open-source software project aimed at providing a platform for rapidly developing robust, easy to use and highly optimised code for TNT calculations. The objectives of this paper are (i) to give an overview of the structure of TNT library, and (ii) to help scientists decide whether to use the TNT library in their research. We show how to employ the TNT routines by giving examples of ground-state and dynamical calculations of one-dimensional bosonic lattice systems. We also discuss different options for gaining access to the software available at \url{http://www.tensornetworktheory.org}. \\

\end{abstract}


\maketitle

\section{Introduction}

Tensor Network Theory (TNT) is a powerful approach to numerically solve problems in physics \cite{Schollwock2011,Verstraete2008,Orus2014,Evenbly2014}, mathematics \cite{Oseledets2011} and computer science \cite{Cichocki2014}. TNT algorithms require optimized software for storing and processing high dimensional complex multi-linear data and interfacing it with standard linear algebra packages. The TNT library provides a unified framework for the efficient implementation of existing TNT algorithms and for the rapid development of custom algorithms.

A common feature of the many varied tensor network geometries and algorithms is that they are built up from a few basic tensor operations. Essentially these boil down to {\em contracting} pairs of tensors to form new ones, {\em reshaping} tensors by combining or splitting legs indices, and correspondingly applying standard linear algebra operations to reshaped tensors that factorise them either via a singular value decomposition (SVD) or an eigenvalue decomposition. Thus on the face of it TNT algorithms are very simple and can easily be described using a graphical representation that we will introduce below. However writing efficient software to perform algorithms quickly becomes complex. The numerous reshapes and re-ordering of the tensors, along with information keeping track of global physical symmetries on the indices \cite{Singh2011} necessitates efficient software to handle these manipulations.

The aim of the TNT project is to provide, from the outset, completely general but highly optimised software for handling these `building blocks' and to allow a broad range of users to benefit from it. To achieve this goal the library consists of three software tiers, with well-defined boundaries between them, that cater for different types of users (c.f. \fir{fig:structure}). Tier 0 interfaces multi-linear data with standard linear algebra libraries and will remain hidden from most users. Tier 1 provides the basic functionality for tensor storage and manipulation. Tier 1 functions exposed to the user are designed to enable rapid development of custom TNT algorithms and tensor networks. These functions are also utilized in Tier 2 libraries that implement the most common TNT networks like e.g. networks corresponding to matrix product states. Finally, Tier 3 of the library provides ready-made implementations of the most common TNT algorithms.

This tiered structure pools the majority of time-critical code into Tier 0 and ensures that performance improvements can immediately be shared by all TNT algorithms built on top of it. The modularity of the library allows easy extension and inclusion of new TNT paradigms into its Tiers 2 and 3. Finally, the algorithm based Tier 3 layer hides all the complexity of tensor manipulations for standard TNT calculations. They can be used with relatively little effort by non-expert users and can even be incorporated into teaching materials.

Naturally these aims are also shared in part by other successful projects that provide high performance libraries based on tensor network methods, such ALPS~\cite{ALPS}, POWDER~\cite{POWDER}, BlockDMRG~\cite{BlockDMRG}, DMRGapplet~\cite{DMRGapplet}, EvoMPS~\cite{Evomps}, DMRG++\cite{DmrgPlusPlus}, SnakeDMRG~\cite{snakedmrg} and simpleDMRG~\cite{simpledmrg}, which focus on simulations of 1D quantum systems using DMRG and other well-known routines for MPS. Since the TNT project is based firmly on the tensor foundations it resembles more closely the philosophy of the iTensor~\cite{iTensor} and Uni10~\cite{Uni10} projects, however there are differences in the approach taken here, specifically in the programming language and scope.

This paper provides an overview of the structure of the TNT library and aims to help researchers deciding whether the TNT library could be useful to them. While Tiers 0 and 1 are completely general we focus here on the library's applications in quantum many-body problems and give an introduction in \secr{sec:QMBP}. This is followed by a brief description of the basics of TNT in \secr{sec:TNTintro} where we also introduce a widely used graphical tensor network notation. In \secr{sec:structure} we give a detailed account of the structure of the TNT project including simple examples of how to implement basic tensor network operations. This is followed by example calculations in \secr{sec:examples} and a discussion of TNT library performance in \secr{sec:performance}. We conclude the paper with brief sections on how to gain access to the software in \secr{sec:accessing} and an outlook on the future development in \secr{sec:conclusion}. A substantial appendix is also included where we describe some of the important core tensor features of the library to help ease potential users into the documentation that is embedded directly into the TNT library and available online at \url{http://www.tensornetworktheory.org}.

\begin{figure*}[ht]
\begin{center}
\includegraphics[scale=0.7]{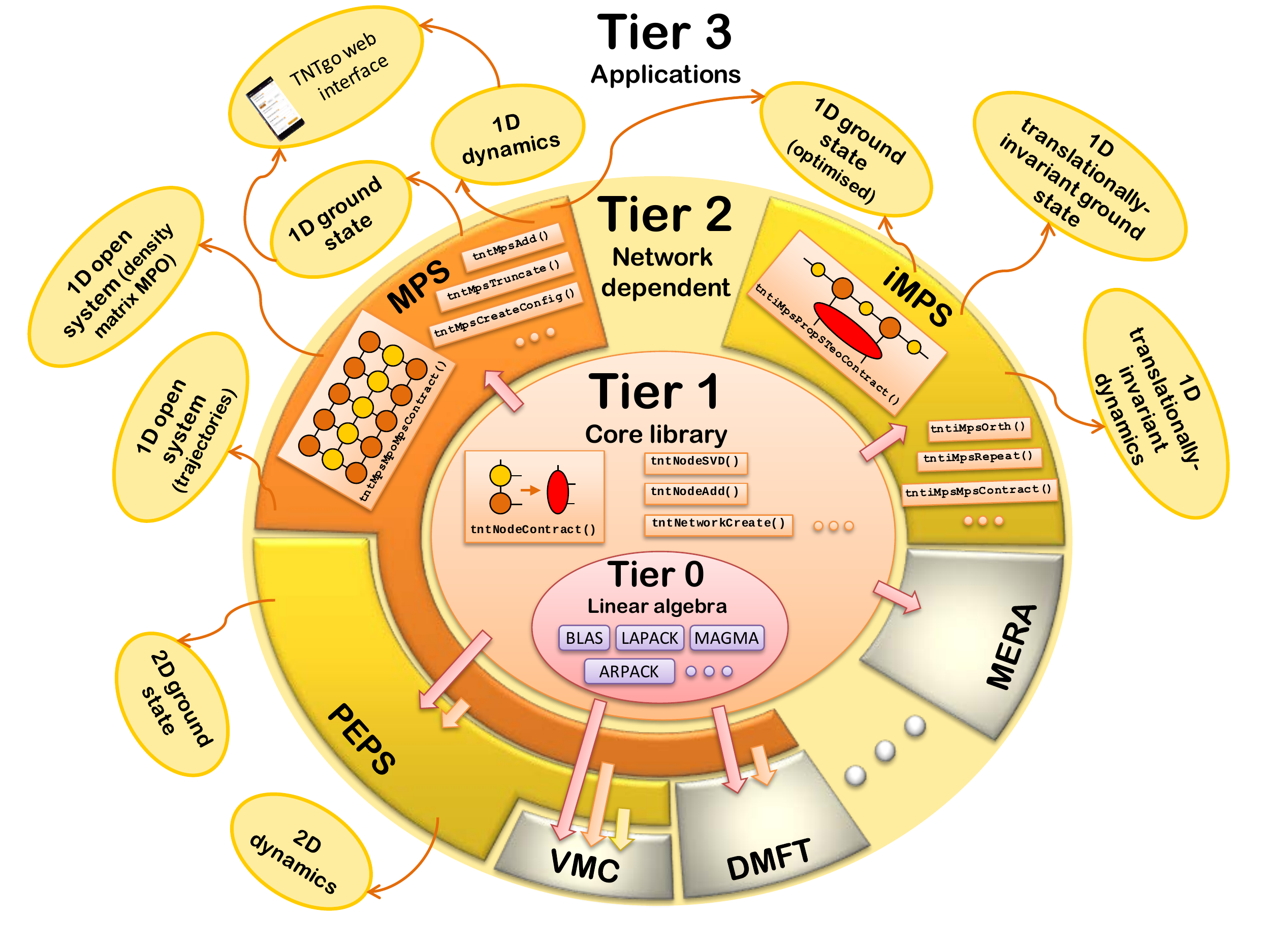}
\end{center}
\caption{Structure of the TNT library. Tier 1 contains functions that do not depend on the network geometry. Tier 2 contains network-specific libraries, all of which are dependent on the Tier 1 core library. Additionally some Tier 2 libraries can be dependent on one another. The infinite matrix product state (iMPS) and projected entangled pair state (PEPS) libraries are still in development, and the variational Monte Carlo (VMC), dynamical mean-field theory (DMFT) and multi-scale renormalisation ansatz (MERA) libraries are planned for future development. Tier 3 contains complete algorithms for performing simulations, which are comprised of Tier 2 `building blocks'.}
\label{fig:structure}
\end{figure*}

\section{The quantum many-body problem}\label{sec:QMBP}
Understanding the collective behaviour of interacting many-body quantum systems remains an outstanding challenge in modern physics. Indeed strong correlation between electrons gives rise to a rich range of phenomena, such as superconductivity, antiferromagnetic ordering, and topological spin liquids, which are both intriguing and functionally relevant \cite{Basov2011}. Further to this in the past decade interest in the coherent many-body dynamics of these systems far-from-equilibrium has dramatically intensified. On the one hand this activity has been propelled by developments in cold-atom experiments \cite{Bloch2008}. These allow for the realisation of controllable, well isolated strongly interacting many-body systems whose evolution can be tracked in real time. On the other hand recent advances in ultrafast THz laser science have now opened up new vistas of experiments probing and controlling solid-state systems \cite{Nicoletti2016}. In particular, selective large amplitude laser excitation of collective modes of a solid may allow for ultrafast switching between different broken-symmetry phases. This not only includes melting equilibrium long-ranged order, such as charge-density waves and superconductivity, but even more remarkably inducing such order with light in regimes where none existed in equilibrium.

\begin{figure}[ht]
\begin{center}
\includegraphics[scale=1.0]{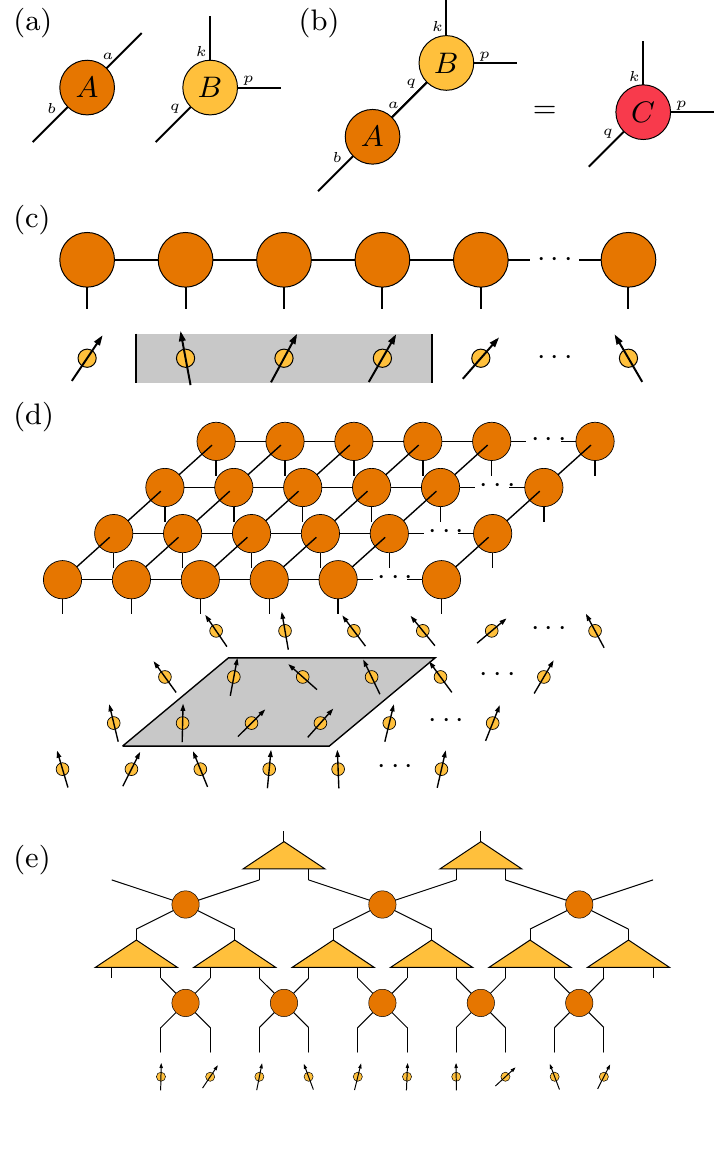}
\end{center}
\caption{(a) A graphical depiction of an order-2 tensor $A$ and an order-3 tensor $B$. (b) The contraction of tensors $A$ and $B$ to form a new order-3 tensor $C$ is shown graphically as the joining of the corresponding legs. (c) An MPS tensor network for 1D systems. The vertical legs are open (uncontracted) and correspond to the physical states of sites of a 1D lattice shown directly beneath. (d) A PEPS tensor network for 2D systems. Again vertical legs map to sites of an underlying lattice. In both (c) and (d) a region of the system is shaded and its boundary with the rest of the system is outlined by the thick lines. (e) A MERA tensor network for 1D systems.}
\label{fig:tntnetworks}
\end{figure}

  A cornerstone of theoretical studies into these systems is the concept of minimal lattice models believed to capture the key physics, e.g. Hubbard-like Hamiltonians defined for one or more relevant electronic bands \cite{Dagotto1994}. Given such a model, understanding phenomena seen in experiments ideally involves solving two key problems: (A) finding the ground state or low-lying excited states of the system, and (B) time-evolving the system from a given initial state according to some quench or periodic driving of the model. Despite their simplicity minimal models are extremely difficult to study; most of them cannot be solved exactly, the regimes of interest cannot be handled perturbatively, and widespread correlations make mean-field methods conceptually inadequate.

  Numerical methods attacking the full quantum many-body problem are therefore essential. However, a fundamental roadblock is encountered - for a system composed of $N$ separate $d$ dimensional degrees of freedom (sites), its quantum state possesses $d^N$ complex amplitudes $\psi_{j_1 j_2\cdots j_N} = \psi_{\bf j}$  defined by $N$ physical indices ${\bf j} = j_1,j_2,\cdots,j_N$. This exponential scaling of the Hilbert space is aptly named the ``curse of dimensionality" and limits exact diagonalisation to very small $N$. Furthermore numerous quantum Monte Carlo methods (e.g. based on imaginary-time projection), which otherwise successfully sidestep this issue for bosons, suffer from the notorious sign problem for crucial fermionic and frustrated lattice systems. Consequently, there exists an acute capability gap between the exciting experiments on strongly correlated systems and the theory trying to unravel their behaviour.

  Over the past 15 years a new approach to the many-body problem has emerged based on Tensor Network Theory (TNT) \cite{Schollwock2011,Verstraete2008,Orus2014,Evenbly2014}. So far it has provided powerful algorithms for solving problems (A) and (B) for 1D and 2D quantum lattice systems by exploiting insights from quantum information science. Central to this is entanglement entropy - a quantity that embodies most generally the notion of quantum correlations in a purse quantum state between two subsystems. Remarkably studies have found that for gapped Hamiltonians with short-ranged interactions the entanglement between two regions in the system's ground state, as well as low-lying excitations, is localized on their shared boundary \cite{Eisert2010}. As a result entanglement of a region scales as a so-called ``area-law", which is a striking contrast with a volume scaling commonly expected for an extensive quantity. This property constrains physical states of minimal models to occupy a very small ``corner" of Hilbert space - a fact that provides a sought-after antidote to the curse of dimensionality. The central aim of TNT is to provide a highly flexible and unifying framework for constructing families of quantum states in this corner. It does so by explicitly designing them to exhibit a given entanglement structure, e.g. like satisfying the area-law.

\section{A brief introduction to Tensor Network Theory}\label{sec:TNTintro}
  The building blocks of TNT are tensors, which in this context are essentially multidimensional arrays of complex numbers. It is useful to visualise them with a diagrammatic formalism where tensors are shapes with legs, each leg associated to a tensor index. In \fir{fig:tntnetworks}(a) an order-two tensor $A_{ab}$, equivalent to a matrix, is shown along with an order-three tensor $B_{pqk}$. If the indices $a$ and $q$ for the tensors $A$ and $B$ have the same dimension then they can be {\em contracted} to give a new order-three tensor $C_{pbk} =\sum_c A_{cb}B_{pck}$ shown in \fir{fig:tntnetworks}(b). Contraction is computationally equivalent to standard matrix multiplication and is diagrammatically represented by joining the corresponding legs of the tensors together. A many-body quantum state $\ket{\psi}$ with complex amplitudes $\psi_{\bf j}$ is therefore an order-$N$ tensor - an intractably large structureless monolithic object.

   To overcome this TNT attempts to factorise $\psi_{\bf j}$ into a network of low order tensors. Specifically we have a network $G$ of vertices $\nu$ each with a tensor $T^{(\nu)}$. These tensors then possess a set of internal indices, of dimension at most $\chi$, and may additionally possess physical indices $j_\nu$. The edges of the network $G$ describe how the internal legs of each tensor are to be joined together and therefore contracted. Thus, in this representation the order-$N$ tensor of amplitudes for a quantum state emerge as the open physical indices left over after performing all the contractions of the internal indices specified by the network $G$, denoted by the tensor trace ${\rm tTr}[\cdots]$ as
\begin{equation}
\psi_{\bf j} = {\rm tTr}[\otimes_{\nu \in G}T^{(\nu)}].
\end{equation}
Important examples are given in \fir{fig:tntnetworks}(c)-(e) and are motivated by dimensionality, e.g. tensor networks mimicking the underlying structure of the physical lattice, or by renormalisation concepts, e.g. multilayered mimicking a Kadanoff-like spin-blocking approach. The matrix product state (MPS) network shown in \fir{fig:tntnetworks}(c) is composed of a chain of tensors contracted together. The thicker open vertical legs are physical indices corresponding to lattice sites of the system shown below the network. Similarly in \fir{fig:tntnetworks}(d) the equivalent projected entangled pair state (PEPS) tensor network is shown that generalises MPS for a 2D square lattice. In \fir{fig:tntnetworks}(e) the multiscale entanglement renormalisation ansatz (MERA) is shown which possesses a hierarchically layered tree network where only the tensors in the bottom layer have physical indices.

The contraction of internal indices between tensors is directly related to the entanglement between the parts of the physical system the tensors are associated with. The more entanglement there is the larger the dimension $\chi$ has to be. If $\chi$ is allowed to scale exponentially with $N$ then in principle a tensor network can describe any state $\ket{\psi}$ with volume scaling entanglement, but suffers the curse of dimensionality again. However, if $\chi$ is bounded the tensor network will only contain polynomially many elements, yet its network geometry can still allow for area-law scaling entanglement within the encoded state. This is illustrated in \fir{fig:tntnetworks}(c)-(d) where a patch of the physical system is shaded. The area-law states that the entanglement of this patch with the rest of the system scales with its boundary outlined by the thick lines. For 1D this is a constant, while in 2D it grows as the perimeter of the patch. In 1D MPS therefore respect the area-law, but do not if they are applied to 2D by e.g. snaking a chain across a lattice \cite{Schollwock2011}. In contrast the higher connectivity of PEPS ensures they continue to obey the area-law in 2D \cite{Verstraete2008}. The MERA network in \fir{fig:tntnetworks}(e) can capture entanglement stratified over many length scales as seen for critical systems that logarithmically violate the area-law \cite{Evenbly2014}.

  To solve problem (A) TNT algorithms employ a variational approach on the class of states described by a given tensor network with limited $\chi$. Several properties set this variational approach apart from others. First, tensor networks provide an enormous class of variational states defined by many hundreds of thousands of parameters, depending on the value of $\chi$. As such increasing $\chi$ can easily refine the ansatz. Second, the bias of tensor networks is extremely general in the sense that it relates to entanglement, as opposed to any specific type of correlation or ordering. Indeed, beyond common spin or charge ordering tensor networks can also readily capture global topological properties and hidden string ordering via local symmetries of the tensors. Moreover this bias weakens quickly with increasing $\chi$. But even more crucially this bias in the entanglement structure is commensurate to that of the corner of Hilbert space occupied by physically relevant states. As such tensor networks are expected to provide highly accurate and enormously compressed representations of physically relevant states and can be considered a quasi-exact variational optimisation.

  To solve (B) a tensor network is used to capture dynamics via a Trotterised update scheme or via the time-dependent variational principle. The success of this relies on the fact that low-lying excited states are also contained in the area-law constrained corner. However, entanglement typically grows rapidly when a system is perturbed strongly requiring that $\chi$ increases to compensate. As a result time-evolving states can often only be adaptively tracked for short timescales.

  For 1D systems MPS form the basis of hugely successful methods such as density matrix renormalisation group (DMRG) \cite{White1992, Schollwock2011}, that solves (A), and time-evolving block decimation (TEBD) \cite{Vidal2003}, that solves (B). Both these algorithms scale as $\chi^3$ and the reachable extreme of $\chi_{\rm MPS} \sim 10,000$, after exploiting SU(2) symmetry, has made a substantial amount of equilibrium physics and dynamics accessible for 1D systems governed by local Hamiltonians \cite{Schollwock2011}. For 2D systems PEPS in principle can mimic the success of MPS in 1D once $\chi \sim 10$ owing to the higher connectivity of the network \cite{Verstraete2008}. However, the higher order of the tensors in PEPS also presents a major computational barrier since algorithms to variationally optimise them scale as $\chi^{10}$. Although polynomial this is nonetheless a formidable scaling that has severely limited practical calculations to an often insufficient $\chi_{\rm PEPS} \sim 5$. The story is similar for MERA where the scaling for algorithms is $\chi^{13}$ \cite{Evenbly2014}. So while MPS based algorithms have succeeded on workstations and turn-key commodity clusters with only mild optimisations, the further development of PEPS and MERA will necessitate the development of code tailored to high performance computing environments. This aim motivates our work on the TNT library.

\section{Structure of the TNT library} \label{sec:structure}
The flexibility of the TNT library is achieved through its tiered structure shown in \fir{fig:structure}. Tiers 0 and 1 comprise the core library, on which all other tiers are based. As such Tier 1  contains tensor routines that are general and do not depend on any aspects of the network or physical system. Tier 2 contains `plug-in' libraries each of which relate to a specific network geometry, and is composed of routines for manipulating these networks. Tier 3 contains complete applications making use of one or more Tier 2 libraries to build an algorithm that accepts the physical parameters of the system as an input and outputs the required results. A complete description of Tier 2 and Tier 3, e.g. for MPS and PEPS algorithms, will be made in separate technical papers. Here we will provide a brief overview illustrating the main methodology and philosophy of the TNT library.

The source code for all tiers is written in C, which was chosen for performance and portability. These performance considerations are most important for the core tensor operations in Tier 0, which contain the computationally heavy processing of the large arrays arising from tensors in the network. Top level operations on networks are not as computationally intensive and so higher-level programming languages can be used without significantly impacting performance. For example Python can be easily interfaced with C and allows for a full object-orientated wrapper to be developed. This will be included in future releases of the TNT library. We now give a brief discussion of each tier.

\subsection{Tier 0: Linear algebra routines}
Users of the library will normally not interface directly with this tier, however all Tier 1 functions that modify tensor values are dependent upon it. These functions are concerned with the heavy-processing tasks of reshaping tensors to form matrices for matrix multiplications and matrix decomposition. These operations are passed to external linear algebra libraries containing optimised algorithms. Indeed the choice of linear algebra library that the compilation links to has a major impact on the performance of the TNT library, which is described further in \secr{sec:performance}.

\subsection{Tier 1: General node routines} \label{sec:tier1}
Tier 1 contains routines for manipulating a single node, a small group of nodes and limited geometry-independent modifications of networks of tensors. It is therefore the first point of entry for writing tensor network algorithms. It includes routines for modifying the tensor values through operations on the nodes, changing how the nodes are connected to one another in the network, getting certain values (e.g. diagonal values) of the tensors and for contracting small groups of tensors together. The routines are contained in the core library \texttt{libtnt}.

To illustrate the usage of Tier 1 we give an example of a user function \code{HeffA\_contract()} in Listing \ref{lis:varmincontract}. The function \code{HeffA\_contract()} performs a specific contraction of a tensor \code{A}, stored as a \code{tntNode}, with a tensor network \code{H\_eff}, stored as a \code{tntNetwork}. The \code{tntNetwork} type is a convenient wrapper for a set of connected tensors forming a linked list with first and last tensors to serve as navigation points for algorithms. The comments in the listing explain how the code works, however more details on the Tier 1 variable types and functions used can be found in \ref{sec:objects} and \ref{sec:functions}.

\begin{lstlisting}[caption={Example of a function that defines a contraction sequence to be used as the matrix-vector multiplication function},label=lis:varmincontract,escapeinside=||]
tntNode HeffA_contract(tntNode A,
                       tntNetwork H_eff)
|\includegraphics[scale=1.0]{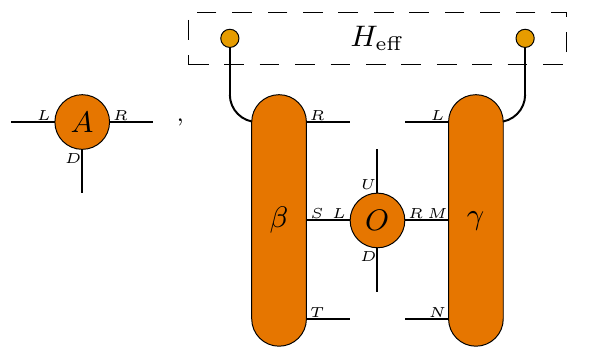}|
{
	tntNetwork H_c; 				// Declare a local network 
    tntNode beta, gamma, O;			// Declare local nodes

    H_c = tntNetworkCopy(H_eff); 	// Make a copy of the input network

    beta = tntNodeFindFirst(H_c);	// Assign beta to first node in H_c
    gamma = tntNodeFindLast(H_c);	// Assign gamma to the last node in H_c 
    O = tntNodeFindConn(beta, "S");	// Assign O to the node connected to beta

    tntNetworkToNodeGroup(&H_c,1);	// Strip away network information 
		|\includegraphics[scale=1.0]{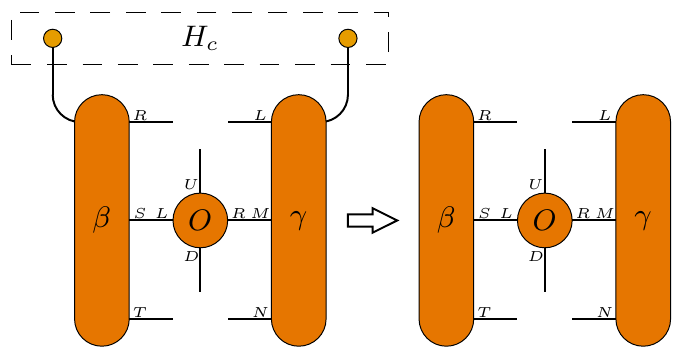}|

    tntNodeJoin(A,"L",beta,"R");	// Join the node A to beta
    tntNodeJoin(A,"R",gamma,"L");	// Join the node A to gamma
    tntNodeJoin(A,"D",O,"U"); 		// Join the node A to O
    // Contract the group of nodes beta, gamma, O and A, the output legs of the result to conform to the convention used on the input A originally
    A = tntNodeListContract("LRD", beta, gamma, O, A);
		|\includegraphics[scale=1.0]{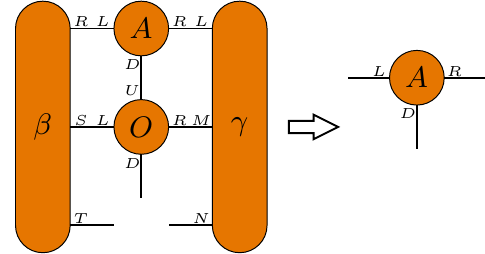}|
    return A;	 // Pass the resulting tensor back
}
\end{lstlisting}

Some linear algebra routines, such as diagonalisation, are iterative and are designed to accept user-defined external functions implementing matrix-vector multiplication, rather than the matrix itself. This allows a problem-specific sparsity structure to be exploited significantly speeding up calculations. In the case of tensor network algorithms such a sparsity structure is usually that the matrix in question is defined by a small network of nodes. Currently there is one Tier 1 routine of this type in the core library wrapped up in \code{tntNetworkMinSite()}, which finds the extremal eigenvectors and corresponding eigenvalues of a matrix. More functions of this type are planned in future releases, for example using reverse-communication when solving large sets of linear equations.

The function \code{tntNetworkMinSite()} expects as an argument a network representing the matrix, a node representing the vector, and a pointer to another function (the contractor) that performs the network contraction corresponding to the matrix-vector multiplication. By taking a contractor as an argument this function is completely geometry independent. The function \code{HeffA\_contract()} given in Listing \ref{lis:varmincontract} is an example of a matrix \code{H\_eff} vector \code{A} multiplying contractor and is useful in MPS ground state calculations, as we shall see shortly.

\subsection{Tier 2: Geometry dependent routines} \label{sec:tier2}
Tier 2 contains routines that are specific to the network geometry and are the building blocks of the TNT algorithms described in \secr{sec:tier3}. All the routines for a specific network geometry are grouped into separate libraries. The library \texttt{libtntmps} contains a suite of routines which act on matrix product states with open boundary conditions. Future releases currently in development include \texttt{libtntimps} which will contain routines acting on infinite MPS systems and \texttt{libtntpeps} which will contain routines acting on two-dimensional PEPS networks. All Tier 2 routines use Tier 1 functions to manipulate the nodes and networks, and thus are dependent on the core library. Additionally there may be dependencies between Tier 2 libraries e.g. PEPS algorithms contain contraction steps that are based on MPS algorithms.

\begin{figure}[ht]
\begin{center}
\includegraphics[scale=1.2]{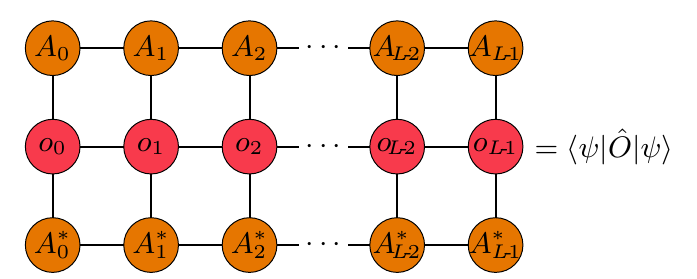}
\end{center}
\caption{This tensor network represents the expectation value of a matrix product operator $\hat O$ of a system in state $\ket{\psi}$. The decomposition of the state into an MPS is usually the task performed by MPS algorithms like TEBD and DMRG. The decomposition of an operator like the Hamiltonian into an MPO can be done analytically, e.g. by hand, straightforwardly.}
\label{fig:mpspmpomps}
\end{figure}

To illustrate usage of Tier 2 we focus on a variational MPS calculation for the ground state of a Hamiltonian. The variational approach relies on minimising the expectation value of the energy. A wide class of Hamiltonians for one-dimensional systems can be exactly formulated as a matrix product operator (MPO), e.g. as a chain-like tensor network. The expectation value $\langle\psi|\hat{O}|\psi\rangle$ for an MPS $\ket{\psi}$ of an MPO $\hat{O}$ is then a contraction of the tensor network shown in \fir{fig:mpspmpomps}. The library \texttt{libtntmps} contains routines for performing such calculations. The variational minimisation of the MPS can then proceed as an alternating local minimisation of each tensor $A$ in the MPS. 

The function \code{ground\_state\_LR()} given in Listing \ref{lis:varmin} performs a very simple, non-optimised one-site update to find the ground state MPS representation for a given Hamiltonian MPO. Some use is made in this function of variable types and functions defined in \texttt{libtntmps} to extract information about the MPS network, however the key steps boil down to Tier 1 operations. 

\begin{lstlisting}[caption={Performing variational minimisation on a group of nodes},label=lis:varmin,escapeinside=||]
double ground_state_LR(tntNetwork psi,
					   tntNetwork H)
|\includegraphics[scale=1.0]{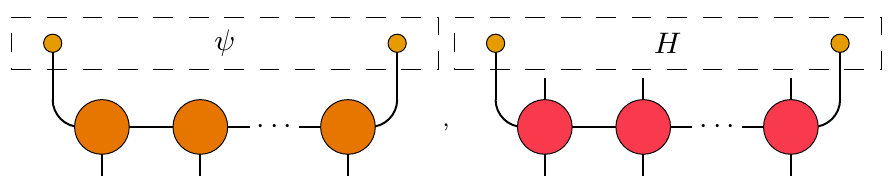}|
{
   tntNetwork psiHpsi, Heff;	// Local networks
   tntNode A_eigv, A;			// Local nodes
   unsigned k, L; 				// Loop and length variables
   double E_eig;				// Energy eigenvalue

   psiHpsi = tntMpsMpoMpsConnect(psi,H); 	// Contract psi and its conjugate with H
   |\includegraphics[scale=1.0]{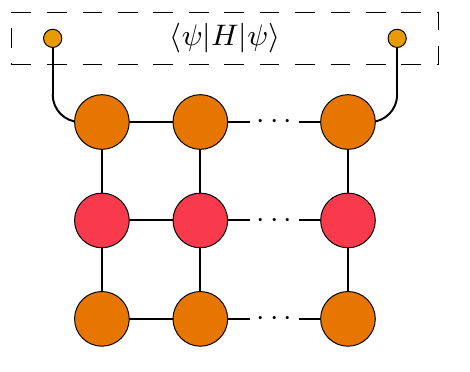}|
   L = tntMpsLength(psi);		// Extract the length of psi
   A = tntNodeFindFirst(psi);	// Assign A to the first node in psi 
   |\includegraphics[scale=1.0]{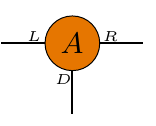}|
   // Loop over each site from left to right
   for (k = 0; k < L; k++) {
        // Contract everything but A on site k to form a local network H_eff
        H_eff = H_prepare(psiHpsi, k); 
        |\includegraphics[scale=1.0]{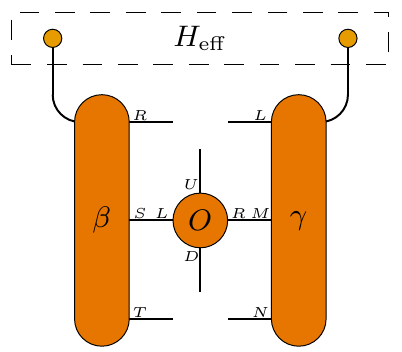}|
        // Perform minimisation of A for this local network H_eff
        A_eig = tntNetworkMinSite(A, &H_eff, 1, &HeffA_contract, NULL, &E_eig);
        // Replace A (linked to psi) with the minimised tensor A_eig 
        tntNodeReplace(A,A_eig); 
       // Move A to point at the next tensor to the right in psi
        A = tntNodeFindConn(A,"R"); 
   }
   return E_eig; // Return the final energy eigenvalue
}
\end{lstlisting}

The function \code{ground\_state\_LR()} sweeps through an MPS network left to right. For each site the network that defines the effective Hamiltonian is prepared (line 18) and the network minimisation routine \code{tntNetworkMinSite()} is then called (line 21). In this example the \code{tntNetworkMinSite()} uses a pointer to the \code{HeffA\_contract()} function given in Listing \ref{lis:varmincontract} and uses the input tensor $A$ as the initial guess for the eigenvector. The \code{tntNetworkMinSite()} routine passes the resulting vector to a suitable iterative sparse eigenvalue solver in an external library. The solver then returns a new vector, which is reshaped to a new node $A$ to be passed to the contract function \code{HeffA\_contract()} until convergence is achieved. The \code{ground\_state\_LR()} then moves to the next site. In real applications many sweeps back and forth are performed.  

\subsection{Tier 3: Applications} \label{sec:tier3}
Tier 3 is the `physics' layer of the library and is intended to contain complete applications for performing simulations e.g.\ for time evolution or for finding the ground state of a given system. The development plan of the TNT library aims to provide a variety of basic applications covering the most popular TNT algorithms. At the time of writing this includes applications for computing low-lying eigenstates, \code{tntGS}, and time-evolution, \code{tntEvolve}, of one-dimensional systems using MPS methods. These applications can be readily used and in principle modified for new settings without detailed knowledge of the specific TNT algorithms.

Applications for one-dimensional open quantum systems described by a Lindblad master equation will also be available soon. This will include both the quantum trajectories approach \cite{Daley2014} and a full `super-operator' approach for the density matrix of the system \cite{Zwolak2004}. The development of applications for two-dimensional quantum systems is a key aim of the project, focusing on PEPS and MERA algorithms as well as Variational Monte Carlo approaches.

\section{Example calculations}\label{sec:examples}
To illustrate the ease of use of the basic MPS applications available we include here two examples. Specifically, we focus on bosons trapped in a one-dimensional lattice described by the Bose-Hubbard Hamiltonian
\begin{equation}
\hat{H} = -J \sum_{j=1}^{L-1}\left(\hat{b}^\dagger_j\hat{b}_{j+1} + \textrm{h.c.}\right) + \frac{U}{2}\sum_{j=1}^L \hat{n}_j(\hat{n}_j - 1) + V \sum_{j=1}^L (j-j_c)^2\hat{n}_j,
\end{equation}
where $\hat{b}^\dagger_j(\hat{b}_j)$ is the bosonic creation (annihilation) operator for site $j$, and $\hat{n}_j = \hat{b}^\dagger_j\hat{b}_j$ is the corresponding number operator. The hopping amplitude is $J$, the on-site repulsive interaction is $U$, the strength of the harmonic trap is $V$ and its centre is site $j_c$.

The basic MPS applications can be interfaced with in multiple ways. The most general way is to specify parameters in an initialisation file. This allows for considerable flexibility in the set up of the system, e.g. allowing site dependent Hamiltonians to be defined or allowing two (or more) species in the system. It also gives the flexibility for the final state saved from a previous calculation to be used as an input for any of the other initialisation file routines. Information about input and output formats supported by the TNT library is discussed in \ref{sec:io}.

Here we will adopt the simplest approach to using the applications, which is to pass all simulation parameters as command line options. This works in the standard way, for example \texttt{--Jb 1} indicates that the hopping term $-\sum_{j=1}^{L-1}\left(\hat{b}^\dagger_j\hat{b}_{j+1} + \textrm{h.c.}\right)$ should be included with a coupling strength of unity. Likewise for other parameters. The observables to be calculated are then specified by options like \texttt{--Ex2bdagb=ap} indicating that the single-particle density matrix $\rho_{ij} =\av{\hat{b}^{\dagger}_i\hat{b}_j}$ is to be calculated for {\em all pairs} of sites $i,j$ in the system.

\begin{figure}[ht]
\begin{center}
\includegraphics[scale=0.65]{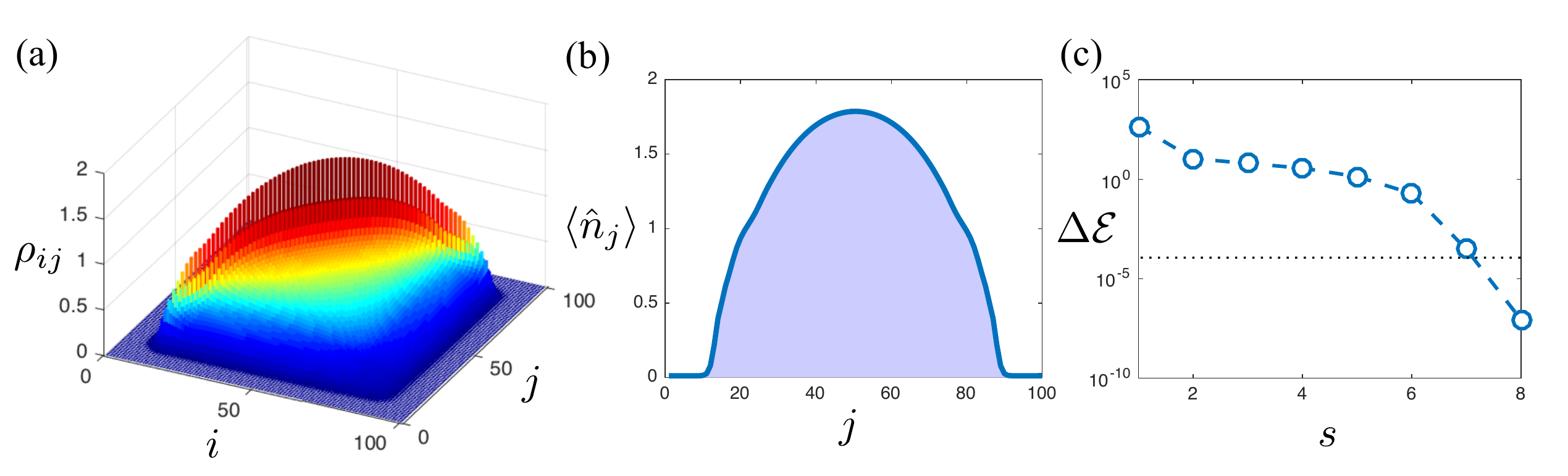}
\end{center}
\caption{(a) The resulting ground state single-particle density matrix $\rho_{ij}$ for the harmonically trapped interacting boson system specified in the main text. (b) The density profile (diagonal of $\rho_{ij}$). (c) The energy difference $\Delta\mathcal{E}$ between successive DMRG iterations $s$ during the calculation. The dotted line indicates the default precision of $10^{-4}$ used by the code, so the code finishes on the 9th iteration.}
\label{fig:ground_state_results}
\end{figure}

In the first example we compute the ground state of interacting bosons for $L=100$ sites in the presence of a harmonic trap at the centre (the default). This is accomplished by the following command line call to \code{tntGS\_cl}:
\begin{mdframed}[style=mdfexample1]
\texttt{./bin/tntGS\_cl -d output/bh\_harm --system=boson --length 100 --n-max 3\\
 -c 100 --qnum-rand-state 100  --Ub 5 --Jb -1 --E-harm 0.01\\
 --Ex1N --Ex2bdagb=ap}
\end{mdframed}
which specifies that exactly 100 bosons populate the system with $U/J = 5$ and $V/J = 0.01$. A $\chi = 100$ is used, along with a bosonic occupancy cut-off of at most 3 bosons per site. The density and single-particle reduced density matrix are saved in the output file. The resulting $\rho_{ij}$ is shown in \fir{fig:ground_state_results}(a) displaying significant off-diagonal correlations consistent with a superfluid state \cite{Clark2004}. The density in \fir{fig:ground_state_results}(b) has a typical inverted parabola shape with kinks around integer filling. The chosen harmonic trapping is sufficient to prevent occupation near the open boundaries. In  \fir{fig:ground_state_results}(c) the energy difference $\Delta\mathcal{E}$ between successive DMRG iterations (similar to Listing \ref{lis:varmin}) is shown and stopped after a given precision is reached (this defaults to $10^{-4}$ but can be set via  \texttt{--precision} option).

\begin{figure}[ht]
\begin{center}
\includegraphics[scale=0.65]{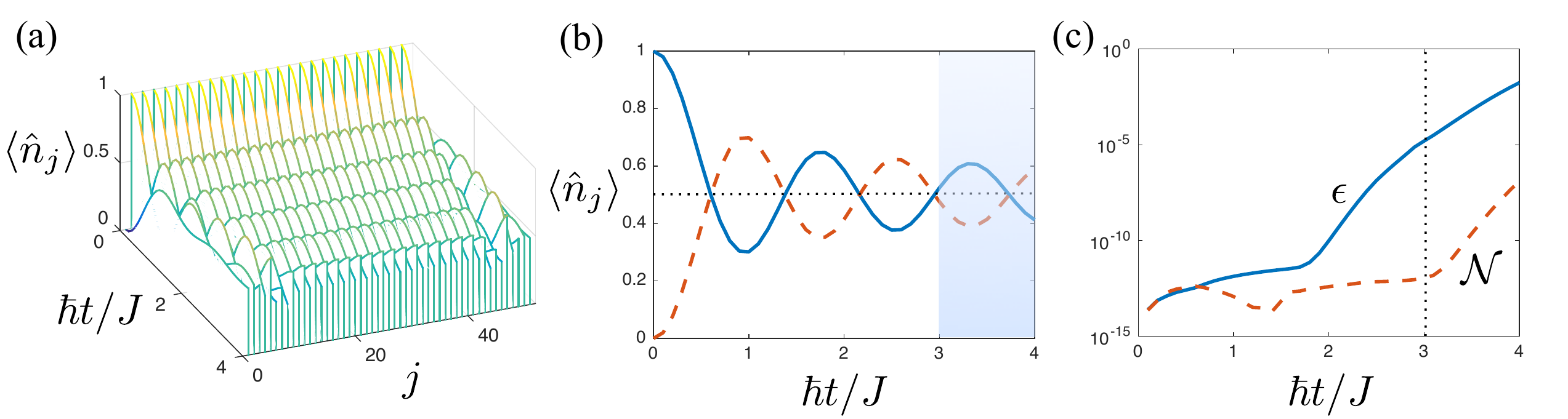}
\end{center}
\caption{(a) The time-evolution of the density $\av{\hat{n}_j}$ across the system starting from a CDW state. (b) The density $\av{\hat{n}_j}$ for the central site $j_c = 26$ (solid) and its neighbour (dashed). The dotted line is the expected asymptotic value $\av{\hat{n}_j} = 0.49$, while the shaded region is where the simulation is no longer accurate. (c) The accumulated truncation error $\epsilon$ and the deviation of the norm $\mathcal{N} = 1 - \braket{\psi}{\psi}$ of the time-evolved state. After $\hbar t/J \approx 3$ both these metrics are becoming unacceptably large.}
\label{fig:time_evolve_results}
\end{figure}

For the second example we compute the time-evolution for a system of hard-core bosons on $L=51$ sites initialised in a `charge density wave' (CDW) state with alternating filled and empty sites and no harmonic trap, i.e. just open boundaries. This is accomplished by the following command line call to \code{tntEvolve\_cl}:
\begin{mdframed}[style=mdfexample1]
\noindent \texttt{./bin/tntEvolve\_cl -d output/bh\_cdw --system=boson --length 51\\
 --n-max 1 --Jb -1 --Ex1N -t 400 -b 10 --dt 0.01 -c 200\\
--qnum-config-state 010101010101010101010101010101010101010101010101010}
\end{mdframed}
which specifies the CDW configuration of 25 bosons. Time-evolution is computed for 400 time-steps each with an increment of $\delta t = 0.01 \hbar/J$ and a $\chi = 200$. Only the density was specified as an observable. This is shown in \fir{fig:time_evolve_results}(a) over the system in time, and in more detail for the central site and its neighbour in \fir{fig:time_evolve_results}(b). As expected the hopping causes this CDW state to melt causing the densities to quickly exhibit small oscillations about the expected asymptotic value of $\av{\hat{n}_j} = 25/51 \approx 0.49$. While this dynamics is easily understood it is in fact quite demanding numerically. In \fir{fig:time_evolve_results}(c) the accumulative truncation error $\epsilon$ (see \eqr{eq:trunc_err} later) grows in time, in particular acutely after $\hbar t/J \geq 2$, and eventually becomes unacceptably large at around $\hbar t/J \approx 3$. At this time the deviation in the normalisation of the time-evolved state $\mathcal{N} = 1 - \braket{\psi}{\psi}$ also displays a large increase. This indicates that with a $\chi = 200$ this simulation can only reliably track the evolution for about 3 hopping times. A similar situation to this example was simulated with much larger $\chi$'s and compared directly to a cold-atom experiment implementing this quantum quench \cite{Trotzky2012}.

\section{Performance} \label{sec:performance}
The design of the Tier 1 \tnode~and \tnet~variables and structures is intended to make developing TNT algorithms as simple as possible, while impacting as little as possible on the performance of these routines. This is achieved by using `opaque' structures for these variables, which not only simplifies using the TNT library, but allows optimisations to be carried out in Tier 0 without any additional steps being taken by users.

As described in \ref{sec:symmnodes} providing symmetry information on legs of the nodes leads to an internal block structure for the tensor. This reduces memory requirements and also means that all linear algebra operations can be carried out block-wise leading to large speed-ups (depending on the physical parameters of the system) - see Tables \ref{tab:dmrgtimes} and \ref{tab:tebdtimes} for examples.

\begin{figure} [ht]
\begin{center}
\includegraphics[scale=0.3]{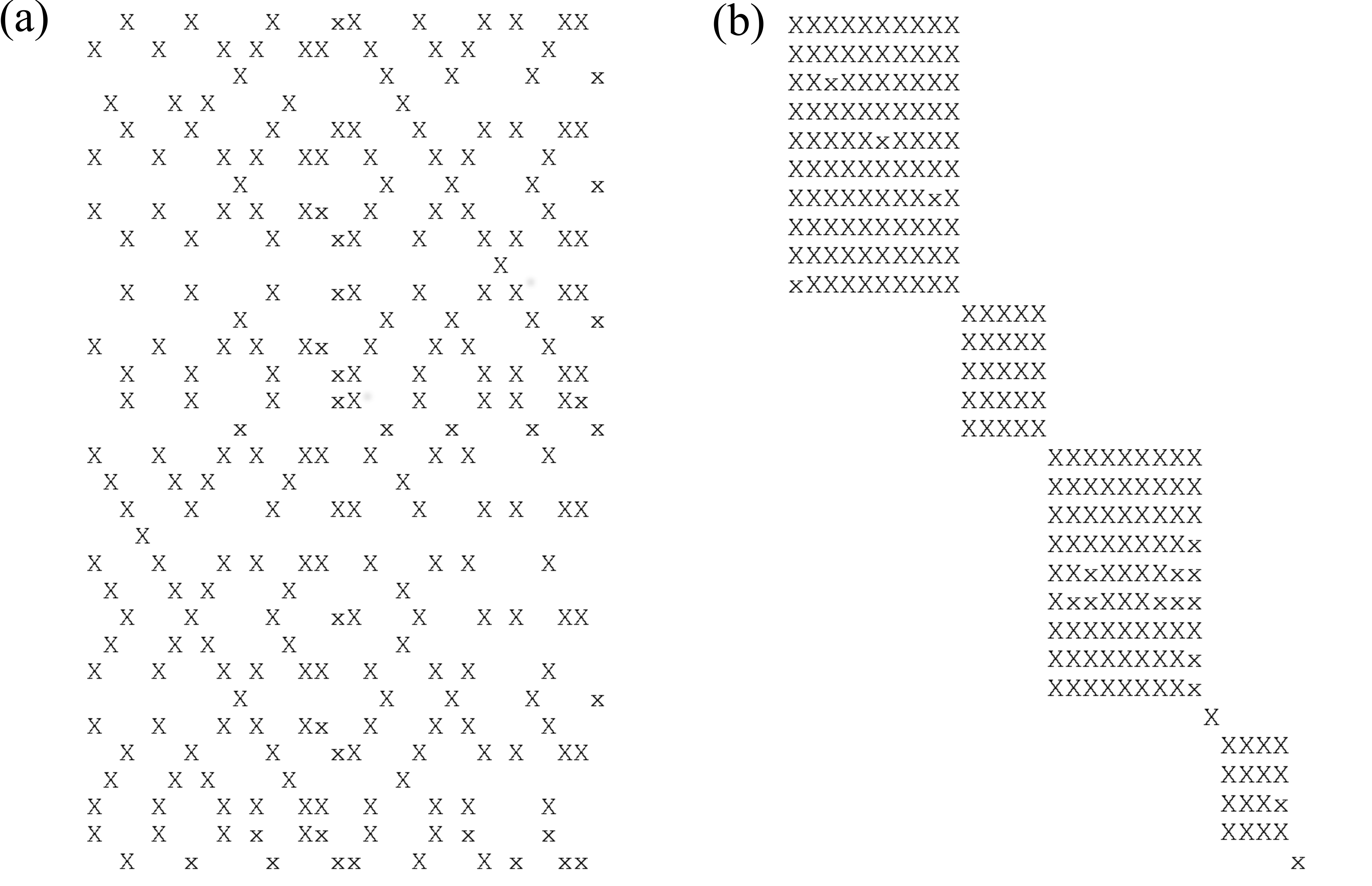}
\end{center}
\caption{Transforming a dense array (a), with entries having magnitude greater than a given tolerance denoted by X, to a blocked matrix (b) by re-ordering the rows and columns.}
\label{fig:autoblocking}
\end{figure}

However, in some cases, the conserved quantities in the system can be difficult to determine, or cannot be assigned directly to the node legs. For these cases, it is possible to enable automatic blocking of the matrices where zeroing all entries smaller than a given tolerance and reordering the rows and columns (see \ref{sec:systemfunctions} for information on setting this) transforms the matrix into a block diagonal form, as shown in \fir{fig:autoblocking}. These much smaller block matrices are then passed to the SVD, and furthermore can be processed in parallel.

\begin{table}
\caption{Performance of the TNT library for 1D ground state simulations. The times are given in seconds for a single DMRG sweep for a spin-1 ($d=3$ physical dimension) isotropic Heisenberg spin-chain of length $L=101$ sites and internal MPS dimension $\chi=200$.}
\label{tab:dmrgtimes}
\begin{indented}
\item[]~\\ \begin{tabular}{@{} cccc}
\br
Blocking type: & none & auto & U(1) \\
\mr
MKL/NAG/MAGMA$^a$ & 617 & 549 & 161 \\
\mr
MKL/NAG$^b$ & 555 & 552 & 135 \\
\mr
MKL/ARPACK$^c$ & 2153 & 2231 & 377 \\
\br
\end{tabular}
~\\
\item[]$^a$ Linking Intel MKL 2015 11.2.2, NAG Fortran library FLL6i25DCL, CUDA 7.5.18, MAGMA 1.7.0 and compiled with Intel compiler 15.0.2.
\item[] $^b$ Linking Intel MKL 2015 11.2.2, NAG Fortran library FLL6i25DCL and compiled with Intel compiler 15.0.2.
\item[] $^c$ Linking Intel MKL 2015 11.2.2, ARPACK-NG 3.3.0 and compiled with Intel compiler 15.0.2.
\item[] $^d$ Linking OpenBLAS 0.2.18, ARPACK-NG 3.3.0 and compiled with GNU compiler 4.9.2 (results not shown).
\end{indented}
\end{table}
Despite the overheads associated with determining the correct row and column order and performing this re-ordering can improve performance since it scales as $O(n^2)$ while the computational complexity of commonly used linear algebra routines like diagonalisation and the SVD scale as $O(n^3)$. However, since the block structure is only realised when performing a linear algebra operation -- the entire matrix is stored for all other operations -- if the global symmetries of the system are known, it is far better to make use of them. Nonetheless, when they are not known turning on automatic blocking can still lead to significant speed improvements as shown in Table \ref{tab:tebdtimes}.

Other optimisations carried out include optional re-use of reshape information when re-ordering indices. This minimises the computation time when tensors of identical dimensions are reshaped in the same way multiple times during a simulation, at the expense of some memory overhead. Determining whether this overhead is worthwhile depends on how many identical reshapes are carried out in a simulation, which will depend largely on whether the internal dimension is predominantly uniform throughout the system.

Care has also been taken to ensure that large amounts of data (which invariably belong to the tensor elements) are not copied or moved in memory unnecessarily, and to make use of shared-memory parallelisation where possible for the Tier 0 operations. Due to these efforts, we have tried to reduce as far as possible the time spent within the TNT routines themselves, so the largest portion of time is spent in external linear algebra routines. When U(1) symmetries are not applied, we find that for large systems as much as 95\% of the CPU time can be spent in linear algebra routines. When U(1) symmetries are used extra processing of the blocks increases the time spent in Tier 1 routines to around half of the total (but much reduced) CPU time.

When choosing an external library it is possible to take advantage of the different computing architectures available. As is now standard, we make use of shared-memory threading with the linked libraries. In addition, CPUs with an integrated GPU are becoming more widespread, and the TNT library can be linked to libraries for these system types. For GPUs we use the MAGMA linear algebra library and have found good performance when the matrix size is large enough (of order a few thousand). The core library routines will automatically determine whether to use GPU or CPU SVD based on the problem size. 

For a comparison of performance with different library types for both ground state and time-evolution calculations see Tables \ref{tab:dmrgtimes} and \ref{tab:tebdtimes}. These simulations were carried out on a single node of the ARCUS-B cluster at the ARC \cite{richardsarc} in Oxford comprising of an Intel E5-2640v3 Haswell 16 core processor with 64GB of RAM and a NVidia K40 GPU. The simulations performed used the Tier 3 initialisation file variants of the applications used in \secr{sec:examples}, \code{tntGS\_if} and \code{tntEvolve\_if}, setup for a spin-1 isotropic Heisenberg spin-chain.  

\begin{table}
\caption{Performance of the TNT library for 1D time evolution simulations. The times are given in seconds for a single TEBD time-step for a spin-1 isotropic Heisenberg spin-chain of length $L=101$ sites perturbed by a central spin-flip from its ground state and evolved with an internal MPS dimension $\chi=1000$.}
\label{tab:tebdtimes}
\begin{indented}
\item[]~\\ \begin{tabular}{cccc}
\br
Blocking type: & none & auto & U(1) \\
\mr
MKL/NAG/MAGMA$^a$ & 977 & 394 & 59 \\
\mr
MKL/NAG$^b$ & 4565 & 516 & 57 \\
\mr
MKL/ARPACK$^c$ & 5612 & 902 & 73 \\
\mr
\end{tabular}
\end{indented}
\end{table}

In general, these times show the importance of choosing a high performance library, although it is worth noting a few points. Firstly, when symmetry information is being used, or $\chi$ is not sufficiently large, the size of the matrices sent to the linear algebra routines is much smaller, and thus using the highly parallelised GPU libraries is of limited benefit. Secondly, although the NAG library provides some optimised routines, the main benefit of this is for the sparse system solvers that are found in ARPACK. For algorithms that do not need to make use of these routines (such as TEBD), linking to the NAG library does not have as great a benefit. Note that these calculations were also carried out using OpenBLAS$^d$ but for the large system sizes used in these tests no results were obtained within a practical computation time.

\section{Access and involvement} \label{sec:accessing}
The quickest and simplest way to try out the TNT library is to use our dedicated online simulation tool TNTgo at \url{http://www.tntgo.org}. Small test calculations can be setup in minutes via a sequence of web forms, as shown in \fir{fig:tntgo}. The inputs are then converted into an initialisation file and executed using the Tier 3 applications \code{tntGS\_if} and \code{tntEvolve\_if} on a small commodity cluster in Oxford. The results are then returned and processed into plots that can be viewed on the site.

To facilitate more advanced use of the library we have created the `virtual box' Linux environment \cite{OVB}, shown in \fir{fig:ovm}. Once the virtual machine is opened, no additional installation steps are required, and TNT applications are pre-installed for a variety of physical systems. For step-by-step instructions on using the virtual machine, go to our CCPForge project page \cite{CCPForge} and register to become a member of the project. The instructions can be found in the \textbf{Docs} section.

\begin{figure} [ht]
\begin{center}
\includegraphics[scale=0.42]{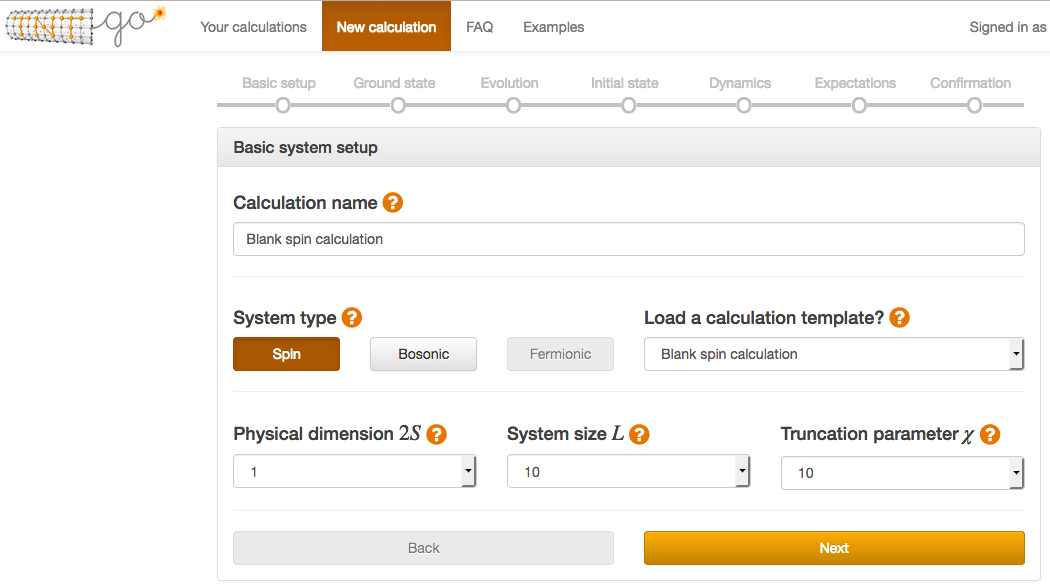}
\end{center}
\caption{A screen shot example of the TNTgo website illustrating how a wide range of 1D calculations can be setup via a sequence of forms. Once completed results from the job can be viewed online.}
\label{fig:tntgo}
\end{figure}

\begin{figure} [ht]
\begin{center}
\includegraphics[scale=0.22]{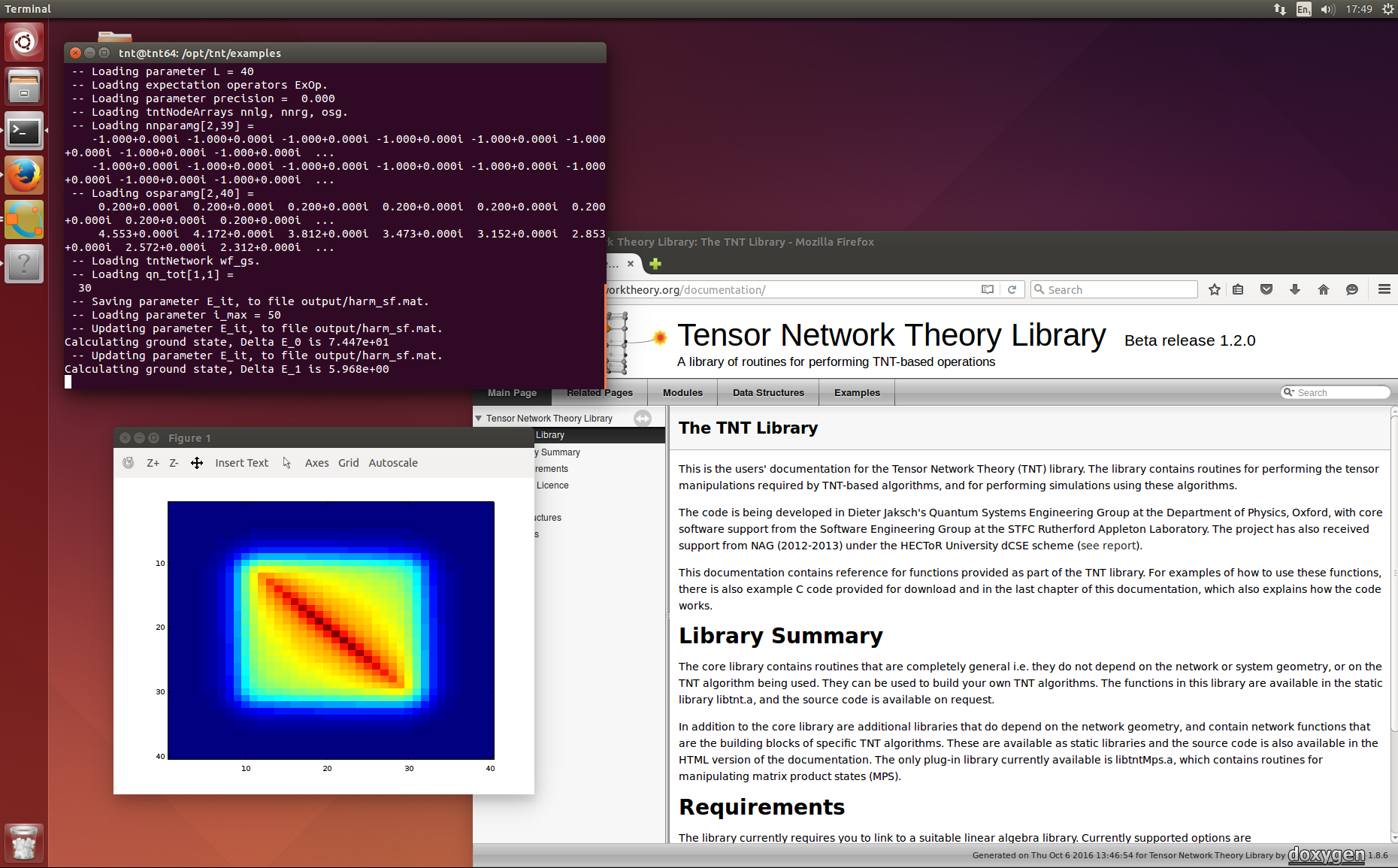}
\end{center}
\caption{A screen shot example of running the TNT virtual machine, which has the TNT library pre-installed on an Ubuntu system. Here a calculation has just been run to find the single particle density matrix of the Bose-Hubbard model in a harmonic trap, like that in \secr{sec:examples}.}
\label{fig:ovm}
\end{figure}

All the routines described in this paper can be downloaded from the \textbf{Releases} section of CCPForge \cite{CCPForge}. The \texttt{libtnt} library is required to run all TNT algorithms. Code for Tier 3 applications described in \secr{sec:tier3} and makefiles which can be modified to run on your own platform are also available for download in the \textbf{Releases} section. Getting the code running requires a C and Fortran compiler (either gcc or icc/ifort), linear algebra libraries (LAPACK, BLAS and either ARPACK or NAG), and at least one of NetCDF or MATLAB libraries for input and output of data. The library \texttt{libtnt} can also be compiled directly from source code, which is the recommended approach for running on high-performance systems. Detailed instructions for this can also be found in the Docs section.

As well as providing a library for TNT simulations, a major aim of the TNT project is to encourage community involvement. This is currently supported by the following features of the CCPForge portal:
\begin{itemize}
\item A discussions forum, for posting information or questions about how to use the routines in the library.
\item A feature requests form, where upcoming features can be viewed, and where anyone can request features they would like to see in future versions of the library.
\item A bug reporting form. Once bugs are fixed you can choose to receive notification of this.
\end{itemize}

It is an eventual aim of the TNT project that users of the library will contribute to routines in Tiers 2 and 3, which will also be handled via CCPForge. Users who wish to contribute to the TNT library can ask for write permission to a development branch of the SVN repository. Modifications can then be incorporated into the latest version of the library. Each branch will be assigned a tracker item (similar to a feature request or branch report) where any notes or discussion of the feature, as well as the differences made to the code, can be viewed. Contributions meeting the required standard will then be integrated in the main line of development code, and included in future official releases.

\section{Summary and future developments}\label{sec:conclusion}
In this technical paper we have presented a high-performance yet easy to use library for performing tensor network simulations. The TNT library is designed to be completely general making it applicable to any type of tensor network. Here we have given an overview of the library structure, discussed examples and analysed its performance for MPS applications. The most important core tensor routines contained in Tier 1 of the library ({\tt libtnt}) are explained in more detail in \ref{sec:objects} and \ref{sec:functions}.

The Tier 2 library ({\tt libtntmps}) for performing MPS simulations in open boundary conditions will be described in detail in an upcoming technical paper similar in style to this one. This will also include applications for open quantum systems and a discussion of the library {\tt libtntimps} tailored for translationally invariant one dimensional systems. Beyond this we envisage further technical papers summarising simulations methods of two-dimensional systems, e.g. {\tt libtntpeps} for PEPS, and more.

\pagebreak

\appendix
\section{Core TNT Variable types} \label{sec:objects}
In this appendix we give a more detailed description of the core TNT library components for Tier 1. The purpose is to give pedagogical examples and explanations to how to use this part of the library, while leaving very detailed information, like function and variable definitions to the online documentation.

All networks in TNT are described using \tnode~and \tnet~variable types. The \tnet~type provides a handle to the entire network, while the \tnode~type defines all the nodes that make up the network. These are defined as so-called `opaque' structures, so that its properties cannot be directly manipulated but are instead accessed through library functions. This ensures backwards compatibility of the code, allowing the flexibility to add or change details of the structures in later versions of the library. This also helps prepare for providing a fully object-oriented interface to the C-library in the future, where the variable types and their associated functions will map to objects and methods within a given class. Additionally there is a global \tsys~variable, which contains details of the simulation and system properties that are shared with all node and network operations.

\subsection{tntNode} \label{sec:node}

Every \tnode~is associated with a tensor i.e.\ a multi-dimensional array of complex numbers. The \tnode~also contains information about how it is connected in the network and symmetric properties of the node.

For the default \tnode~type, the underlying tensor is a simple one-dimensional array of either real or complex numbers in memory, which represents a flattened multidimensional array. Information about the dimensions of and order of the indices is also stored. The tensors can also have additional structure as described below.

Each \tnode~has one or more legs that map onto the indices of the underlying tensor. Each leg will initially map directly to a tensor index, although legs can be fused to create `fat' legs that map onto more than one tensor index. Each leg of a \tnode~can be connected to the leg of the same or another \tnode~to form a network. Legs that are connected represent tensor indices that will eventually be contracted and so only legs with the same dimension can be connected to each other. The legs are labelled and addressed by a single alphanumeric character.

A tensor can be associated with multiple nodes. When a node is copied the (usually large) tensor is simply linked to a new \tnode~object, and so copying nodes is a cheap operation.

\subsubsection{Symmetric nodes} \label{sec:symmnodes}

Symmetric nodes are formed when there is a global physical symmetry in the system. In these cases conservation of the relevant quantum numbers requires that some elements of the tensor are always zero. By ordering the tensor indices suitably the non-zero elements of the tensor can be stored as a group of blocks, each block possessing a given quantum number.

\begin{figure}[ht]
\begin{center}
\includegraphics[scale=1.2]{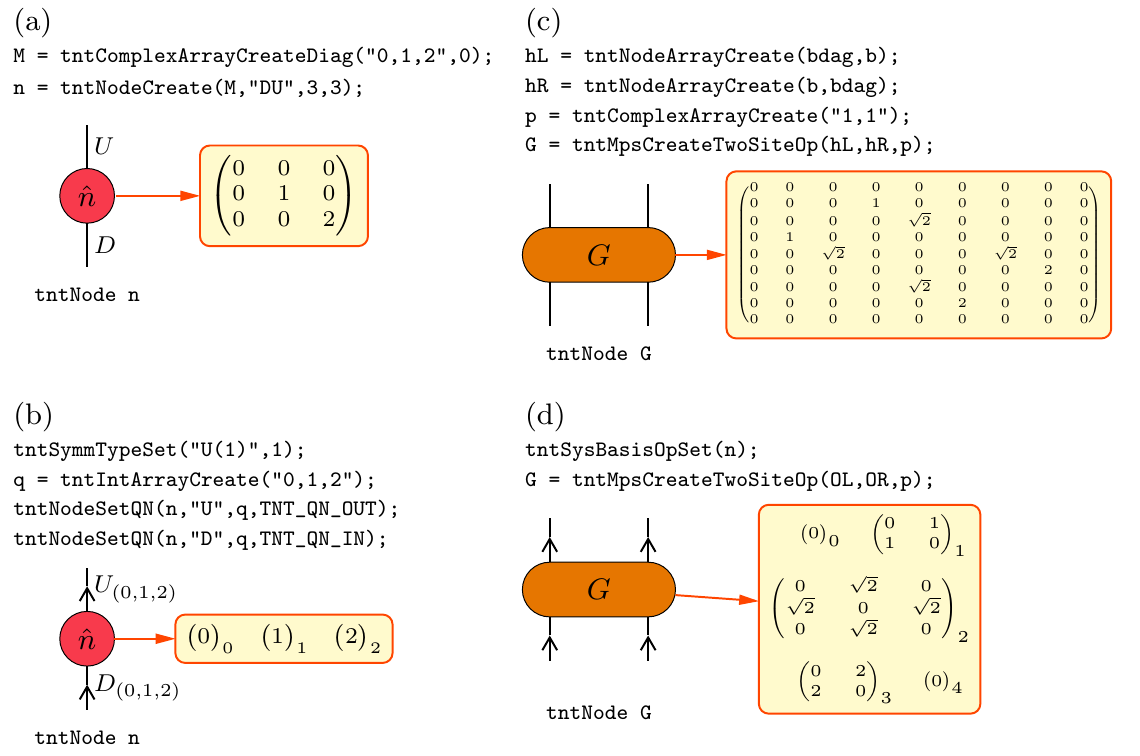}
\end{center}
\caption{(a) The number operator for a bosonic system with no symmetry information. (b) After setting quantum numbers (which represent the possible number of bosons) on all legs the resulting non-zero blocks are simply $1 \times 1$ matrices for each allowed bosonic population on a single site up to a truncation (denoted by the subscript on the matrices). (c) Creating a two-site operator representing the hopping term $\hat{b}^{\dagger}_j\hat{b}_{j+1} + \hat{b}_j\hat{b}^{\dagger}_{j+1}$
without any quantum number information. (d) After the basis operator with symmetries has been set, creating a two-site operator results in a tensor having block-wise form. Here the quantum number of each block represents the number of bosons in the two-site basis of the incoming or outgoing physical legs.}
\label{fig:invariantnodes}
\end{figure}

The TNT library currently supports encoding of U(1) symmetries (e.g.\ for particle number conservation) on nodes. We follow the approach taken in Ref.~\cite{Singh2011}, which achieves this by associating the relevant quantum numbers with each node. Briefly, each \tnode~leg is marked as an incoming or outgoing leg, and every index on a \tnode~leg corresponds to a quantum number label. Once all the quantum numbers on all the legs are set, only tensor elements for which the sum of the incoming quantum numbers equals the sum of the outgoing quantum numbers are retained, and are stored in a block labelled by that sum, as shown in \fir{fig:invariantnodes}. Making use of symmetries not only decreases the memory requirements but also substantially increases computation speed. This is because all tensor operations can be carried out for each quantum number sector, thus decomposing one large tensor operation into several much smaller tensor operations.

The quantum number labels are formed from one or more quantum numbers. For the simple case of conservation of particle number in a single-species system, each quantum number label is a single integer that corresponds to a given number of particles. In an $m$-species system each quantum number label is formed of $m$ integers. This is illustrated in the code listing below for a two-species systems, where the first species $a$ can have 0 or 1 particles per site, and the second species $b$ can have  0, 1 or 2 particles per site.

\begin{lstlisting}[caption={Setting symmetries in a two-species system},label=lis:qnum]
tntIntArray qn;
tntNode na;
tntComplexArray M;

tntSymmTypeSet("U(1)",2);

qn = tntIntArrayCreate("0,0,0,1,1,1;0,1,2,0,1,2");

M = tntComplexArrayCreateDiag("0,0,0,1,1,1",0);
na = tntNodeCreate(&M,"UD",6,6);

tntNodeSetQN(na,"U",qn,TNT_QN_IN);
tntNodeSetQN(nb,"D",qn,TNT_QN_OUT);
\end{lstlisting}

On line 5, symmetries are turned on, where the second argument specifies that there are two quantum numbers per label. Lines 7 and 8 create a two-dimensional array to hold the quantum number labels. The first row lists the quantum numbers for species $a$ and the second row lists the quantum numbers for species $b$. The array thus defines the particle numbers in the single-site basis $\{0_a0_b,0_a1_b,0_a2_b,1_a0_b,1_a1_b,1_a2_b\}$. Lines 10 and 11 create a node that corresponds to the number operator for species $a$, and the quantum numbers are assigned to it in lines 13 and 14.

Care must be taken to ensure that sufficient \tnode~legs (and therefore tensor indices) are defined to allow an invariant tensor to be formed, otherwise elements will be discarded, and a warning is outputted. As shown in \fir{fig:covariantnodes} setting quantum number labels on the physical legs of an operator that results in a change of the total quantum number would cause all the elements to be discarded. However a covariant operator -- one that changes a state from having a well-defined total quantum number label $Q_1$ to one having a different but still well-defined quantum number label $Q_2$ -- can always be reformed as an invariant tensor by adding a singleton leg. Library functions are provided to determine such a suitable quantum number label for the additional leg.

\begin{figure}[ht]
\begin{center}
\includegraphics[scale=1.2]{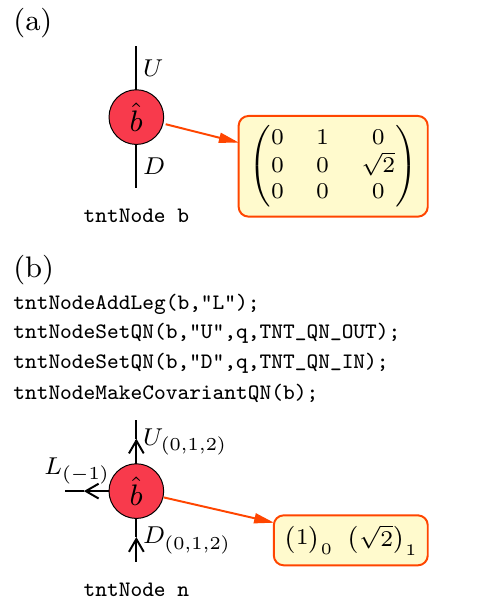}
\end{center}
\caption{Changing a node to a symmetry preserving form. (a) Ladder operators such as $\hat{b}$ shown here change the total quantum number so are not symmetry invariant. (b) However they can be made invariant by adding a singleton leg carrying the appropriate quantum number, and this quantum number can be determined by calling a library function.}
\label{fig:covariantnodes}
\end{figure}

The framework we use to define U(1) symmetries can be straightforwardly and easily extended to other Abelian symmetries, e.g. $Z_q$, within the library by updating the rules for adding quantum numbers on different indices in Tier 0. In the future support for non-Abelian symmetries SU(2) \cite{Singh2012} will be added to the library. These symmetries are encoded in a similar way to Abelian symmetries, e.g. for the case of conservation of total spin the quantum number labels to assign would be composed of two numbers forming the spin index $(j,m)$. The difference between these two symmetry types is contained in Tier 0, where more complicated fusion rules are used when combining indices to reshape the tensors.

\subsubsection{Functional nodes}

A functional node is useful when the tensor elements represented by the \tnode~depend on parameters that change often during a simulation, and are particularly useful when the parameters depend on values that are determined during the calculation itself. Rather than having static values, a functional node is defined by operators and a generating function that are fixed at the time it is created, and parameters that can be changed at any time using library functions.

Functional nodes can be loaded from an initialisation file (defined using a MATLAB library function -- see \secr{sec:io} for more information about input and output). Alternatively, they can be created using core library functions as shown in Listing \ref{lis:funcnode}. First, the single site spin-$\frac{1}{2}$ operators are created from matrices in lines 6 to 12. Each operator is then contracted with itself to form a two-site operator (lines 14, 18 and 22), which are scaled by the time-step (lines 15, 19 and 23), and the matrices for these operators extracted and used to fill an array (lines 16, 20 and 24). The node is then created as a function of the two-site spin coupling terms on line 26. A functional node $F$ can currently either have the form $F = \exp\{\sum_i p_i o_i\}$ or $F = \sum_i p_i o_i$, where $p_i$ are the parameters, and $o_i$ are the operators.  Initially all the parameters are zero. The parameters are then set to form an operator representing propagation under the XYZ Hamiltonian in lines 29 to 31.

\begin{lstlisting}[caption={Creating and setting parameters for a functional node},label=lis:funcnode,escapeinside=||]
	tntComplexArray Mx, My, Mz;
    tntNode Sx, Sy, Sz, O, G;
    tntComplexArray Mo[3];
    tntComplex dt = {0, 0.01};

    Mx = tntComplexArrayCreate("0 1; 1 0");
    My = tntComplexArrayCreate("0 -1i; 1i 0");
    Mz = tntComplexArrayCreateDiag("1 -1", 0);

    Sx = tntNodeCreate(&Mx, "DU", 2, 2);
    Sy = tntNodeCreate(&My, "DU", 2, 2);
    Sz = tntNodeCreate(&Mz, "DU", 2, 2);

    O = tntNodeContract(Sx, Sx, NULL, "DU=EV");
    tntNodeScaleComplex(O,dt);
    Mo[0] = tntNodeGetMatrix(O, "DE", "UV");

    O = tntNodeContract(Sy, Sy, NULL, "DU=EV");
    tntNodeScaleComplex(O,dt);
    Mo[1] = tntNodeGetMatrix(O, "DE", "UV");

    O = tntNodeContract(Sz, Sz, NULL, "DU=EV");
    tntNodeScaleComplex(O,dt);
    Mo[2] = tntNodeGetMatrix(O, "DE", "UV");

    G = tntNodeFuncCreate(Mo, 3, "exp", "DEUV", 2, 2, 2, 2);
    |\includegraphics[scale=1.0]{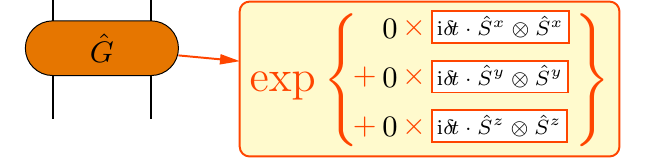}|

    tntNodeSetRealParam(G,1.1,0);
    tntNodeSetRealParam(G,1.2,1);
    tntNodeSetRealParam(G,2.3,2);

    |\includegraphics[scale=1.0]{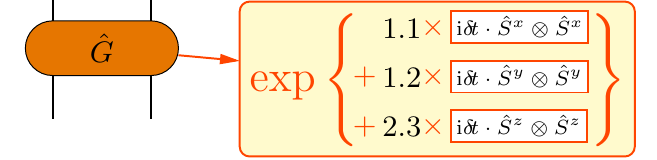}|

\end{lstlisting}

\subsection{tntNetwork}
While small networks of tensors can be described as groups of nodes navigation within an algorithm is made easier by the \code{tntNetwork} type. This provides a convenient handle to a number of nodes that are joined to one another. A network structure is formed of information about the start and end of the network. This is in the form of singleton terminating nodes that are connected to the first and last nodes in the network, represented as the small orange dots in \fir{fig:linkedlist}. It does not contain a list of all the nodes in the network, but instead the network is defined as a linked list as shown in \fir{fig:linkedlist}. This choice makes it very straightforward to insert, remove, contract, or factorise nodes in the network, since only adjacent nodes require updating. Furthermore many algorithms sweep through the network in a sequential pattern, for which a linked list is well suited. To get a handle to any node in the network, the first (or last) node in the network is found and the network then traversed using connections on the legs of the network.

\begin{figure} [ht]
\begin{center}
\includegraphics[scale=1.1]{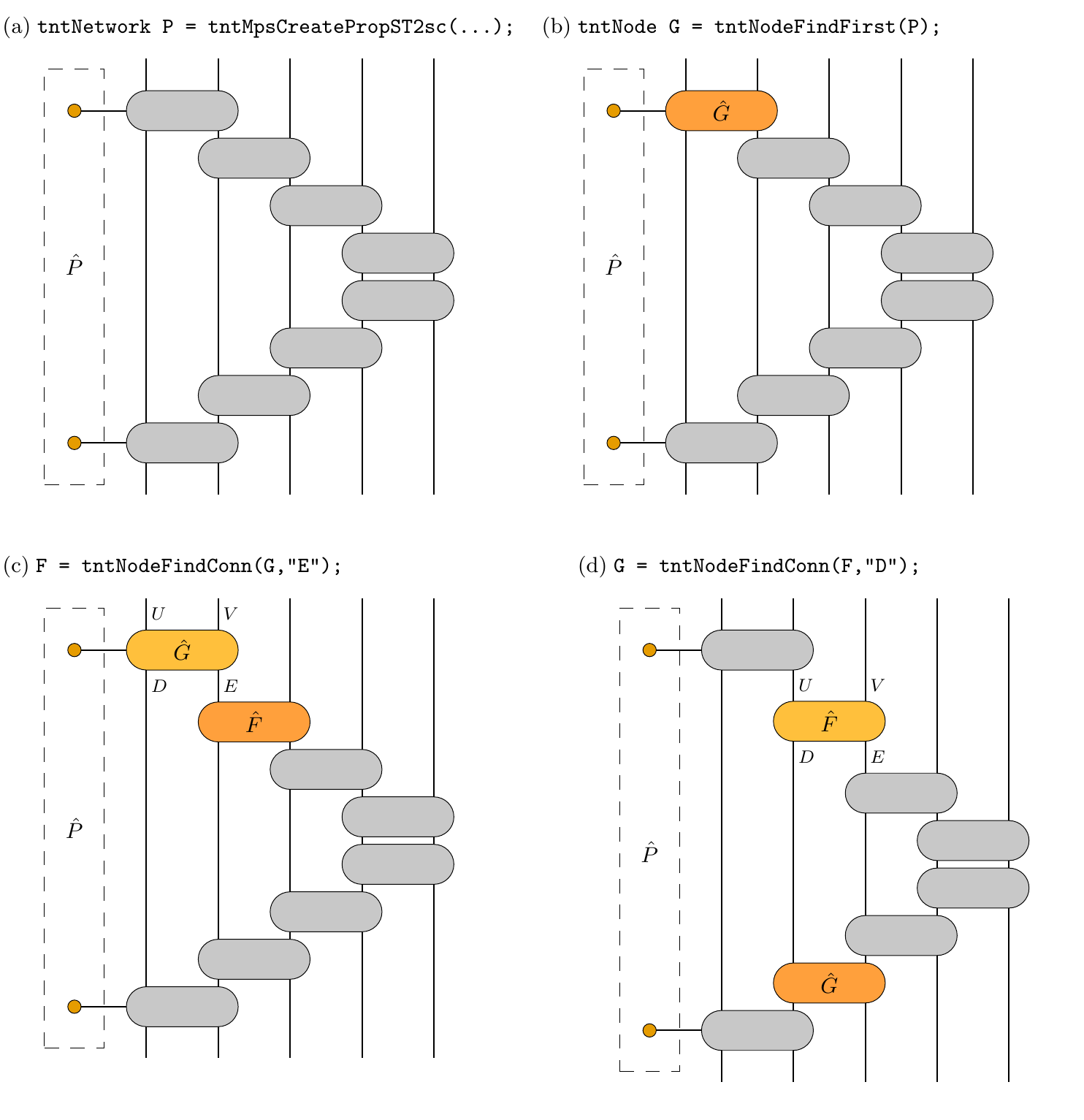}
\end{center}
\caption{Example of creating and traversing a network - items in colour represent those that have a variable handle. (a) The network is created and the variable \texttt{P} contains information about the start and end points of the network. (b) Finding the first node in the network. (c) and (d) Finding other nodes in the network through connections to the currently selected node.}
\label{fig:linkedlist}
\end{figure}

A network can optionally also contain additional (geometry dependent) information required for Tier 2 code. For example an MPS network can contain information about the coefficients for a Schmidt decomposition between each pair of sites in the network. Since this information is always obtained during MPS sweeping algorithms, keeping track of it does not add any extra computation cost.

\subsection{tntSystem} \label{sec:system}
The \tsys~is a global variable that contains a handle to the basis operator \tnode, which is used when generating any networks or nodes that possess physical indices. The basis operator defines the physical basis for the system, for example in a bosonic system the basis operator would be the number operator $\hat{n}$. If there is a global physical symmetry it also contains the quantum numbers for each element of the physical index. Once the basis operator has been defined, any nodes created using library functions that are known to have physical legs can be generated in symmetric form automatically e.g.\ creating starting wave functions or network operators.

As well as this the \tsys~variable holds simulation parameters relating to the SVD type and truncation (see \secr{sec:svd}), the system type (bosonic, fermionic, spin), and tunable parameters for the linear algebra routines (e.g. the maximum number of iterations in the sparse eigenvalue solver). These are first set to their default initial values, however they can be changed by calling core library functions (see \secr{sec:systemfunctions}) or via command line functions. When this is done, a summary of the system information will be output to screen, a example of which is given in the listing below.

\begin{lstlisting}[caption={Printing out system information},label=lis:sys]
------------- System Information -------------
No symmetry type set.
No basis operator set.
Relative truncation tolerance is 1e-16.
Absolute truncation tolerance is -1.
Truncation error tolerance is 1e-08.
Tolerance for automatic blocking is -1.
Current truncation type is 2norm.
SVD type is LAPACK divide and conquer.
Reshape re-use is turned on.
Maximum number of iterations for eigenvalue solver is 300.
\end{lstlisting}

\section{Core library functions} \label{sec:functions}
A full description of all core library functions is given in the TNT documentation \cite{TNTdocs}. Here the most useful and commonly used functions are described.

\subsection{Node algebra}
All the functions which change the values of a tensor associated to a node are based on linear algebra routines, the majority of which call routines in the linked external libraries. These require the tensors to be reshaped into matrices or vectors first by re-ordering the indices appropriately, although this reshaping step is carried out in Tier 0 and so it does not require user input. In some cases the user must specify which \tnode~legs belong to the rows or columns of a matrix, and this is done simply by passing a string of the leg labels -- the remaining manipulation is carried out automatically. This will be explained further by means of some examples below.
\subsubsection{Contraction}
Contracting two tensors together involves performing a sum over tensor indices. For example, consider an order-4 tensor $A_{ijkl}$ and an order-3 tensor $B_{kmn}$: an order-5 tensor $C$ can be formed by summing over the common index $k$
\[
C_{ijlmn} = \sum_k A_{ijkl}B_{kmn}.
\]
In practice, instead of performing a sum over tensor indices, the indices of $A$ and $B$ are reordered, such that the uncontracted indices appear first and last respectively. The contracted indices are assigned to the columns of $A$ and the rows of $B$, and then the tensor contraction simply becomes equivalent to a matrix multiplication $AB$ for which performance threaded linear algebra libraries can be used.

A user of the library performing this contraction would connect \tnode~$A$ to \tnode~$B$ in a network with the legs that correspond to the index $k$ -- let us label these legs $R$ and $L$ respectively -- then call the function to contract them. The reshaping is carried out automatically. Note that since leg labels are unique a \textit{leg map} is required to relabel any remaining legs coming from $A$ and $B$ that have the same label. This is illustrated in \fir{fig:contractpair}.

\begin{figure} [ht] \label{fig:contract}
\begin{center}
\includegraphics[scale=1.2]{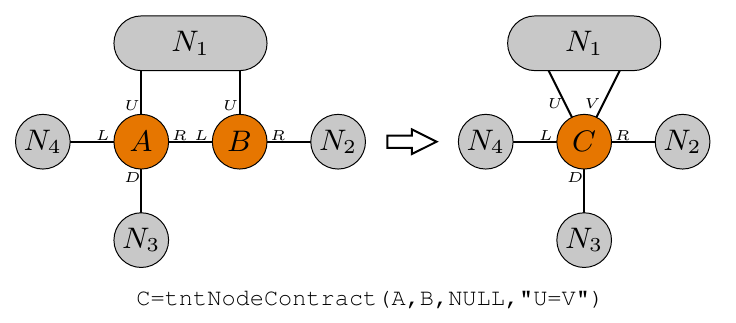}
\end{center}
\caption{Contracting a pair of nodes within a network. Since both $A$ and $B$ have an uncontracted leg labelled $U$, the function call maps leg $U$ of $B$ to a new label $V$ so that all the leg labels are unique on the new tensor $C$.}
\label{fig:contractpair}
\end{figure}

Many algorithms rely on contraction of a whole sequence of nodes, and it is important to do this in an order that minimises the computational cost \cite{Pfeifer2014}. This can be done using a function that performs contraction of a list of nodes. For three or four nodes the contraction cost is explicitly calculated for all permutations of contraction order, and the optimal order chosen automatically. For more than four nodes, the nodes will be contracted in the order they are supplied, with only very minor optimisations. Namely: (a) any legs connected to legs on the same node (which is equivalent to a partial trace) are contracted first; (b) and connections of singleton dimension (tensor product) are always contracted last. An example of contracting multiple nodes for a common contraction that is part of the DMRG algorithm is shown in \fir{fig:contractdmrg}.

\begin{figure} [ht]
\begin{center}
\includegraphics[scale=1.2]{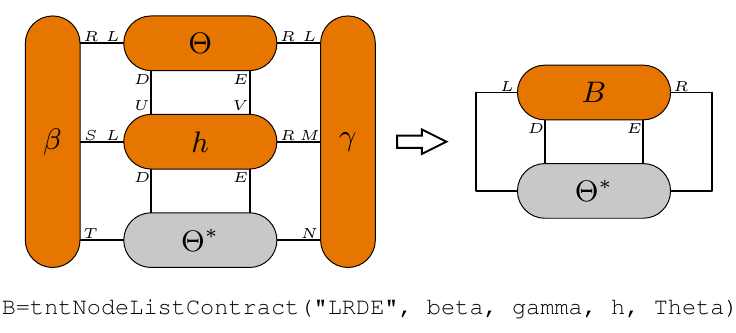}
\end{center}
\caption{Contracting a group of nodes. The first argument of the function lists the new leg labels where the ordering of the legs is given by the order that the nodes that these legs originally belong to appear in the list. In this case the remaining legs after the contraction would be leg $T$ of $\beta$, leg $N$ of $\gamma$ and legs $D$ and $E$ of $h$, which are relabelled as $L$, $R$, $D$ and $E$ on the resulting node $B$.}
\label{fig:contractdmrg}
\end{figure}

\subsubsection{Singular value decomposition} \label{sec:svd}
An SVD factorises a matrix $A$ into a product of three matrices $USV^{\dagger}$, where $U$ and $V$ are unitary matrices, and $S$ is a real diagonal matrix containing singular values $\lambda_i$ listed in decreasing order. To perform an SVD of a tensor having multiple indices, it must first be reshaped to a matrix by assigning the indices to either the row dimension or the column dimension.  The row indices are then assigned to $U$ and the column indices are assigned to $V^{\dagger}$.

A user performing an SVD of a node $A$ into three new nodes $U$, $S$, and $V^{\dagger}$ would call \texttt{tntNodeSVD()} and simply list the labels for the legs of $A$ that correspond to the rows as an argument. In addition the leg labels for the internal legs of $U$ and $V$ and both legs of $S$ should be given. Optionally, the legs can be relabelled after performing the SVD by supplying a leg map. This is illustrated in \fir{fig:svdfig}.

\begin{figure} [ht]
\begin{center}
\includegraphics[scale=1.2]{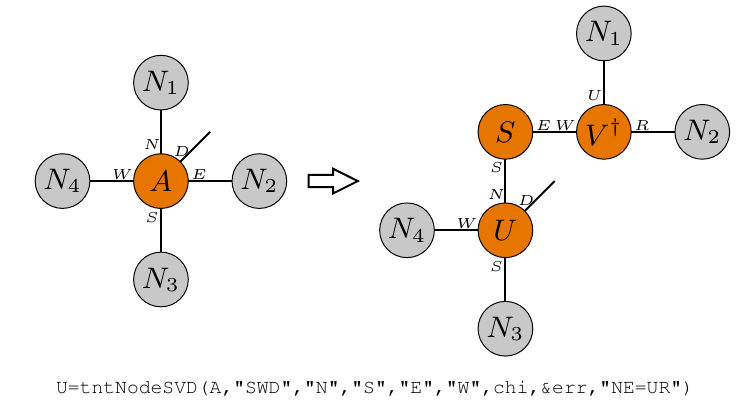}
\end{center}
\caption{Performing an SVD on a multi-legged node connected to other nodes in a network.}
\label{fig:svdfig}
\end{figure}

If an exact SVD is performed, the dimension of the internal legs $D_{\mathrm{exact}}$ will be the minimum of the combined dimension of the legs assigned to $U$ and $V^{\dagger}$. However in tensor network algorithms it is typical to truncate the internal dimension $\chi$ to a value less than this. In all cases there will be an associated truncation error, which by default is calculated as
\begin{equation}
\epsilon_{\mathrm{trunc}} = \sqrt{\left(\sum_{i>\chi} \lambda_i^2\right)}. \label{eq:trunc_err}
\end{equation}
The function used to calculate the truncation error can be changed easily if required (e.g. to the sum of the squares or the 1-norm of the discarded values).

Users of the library can perform a truncated SVD in four ways:
\begin{enumerate}
\item By passing a value of $\chi$ to \code{tntNodeSVD()} that is less than $D_{\mathrm{exact}}$. This will discard all singular values $\lambda_{i>\chi}$.
\item By setting a global absolute truncation tolerance $a$. This will discard all singular values for which $\lambda_{i} < a$.
\item By setting a global relative truncation tolerance $r$. This will discard all singular values for which $\lambda_{i}/\lambda_0 < r$.
\item By setting a global truncation error tolerance $\epsilon_{\mathrm{tol}}$. This will discard the maximum number of singular values for which $\epsilon_{\mathrm{trunc}} < \epsilon_{\mathrm{tol}}$.
\end{enumerate}

Note that when singular values in $S$ are discarded, the associated singular vectors in $U$ and $V$ are also discarded. If two or more of the above bounds are used, then the one which results in the smallest internal dimension $\chi$ will be applied. The choice of the truncation error function and truncation error tolerances are stored in the global \tsys~variable.

\subsubsection{Addition}
Addition of nodes is carried out using the element-wise add function \code{tntNodeAdd()}, as shown in \fir{fig:nodeadd}. Only nodes with an identical structure, i.e.\ the same number, labelling, and dimension of legs, can be added together using this function. If the symmetry information is identical, then addition will be carried out block by block.

\begin{figure}[ht]
\begin{center}
\includegraphics[scale=1.2]{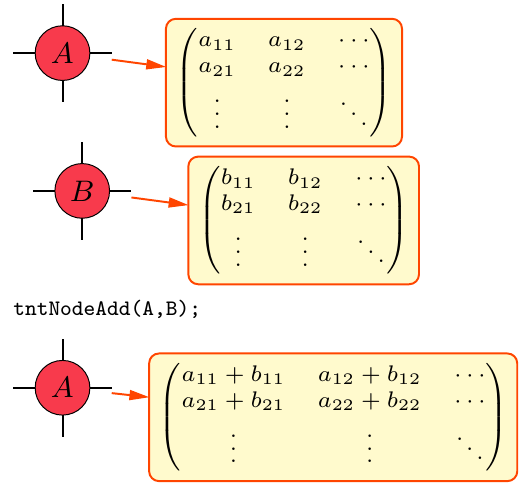}
\end{center}
\caption{Adding two nodes together -- calling the function adds each element of $B$ to the respective element of $A$.}
\label{fig:nodeadd}
\end{figure}

In some algorithms, a direct-sum or tensor-add is required, for example when adding two wave functions to one another by adding their MPS network representations. In this case, the leg dimensions of the nodes must first be expanded, although this may not be on all legs/indices. This is performed using the function \code{tntNodeDirectSum()}. Such operations are crucial to certain TNT algorithms, for example strictly single-site DMRG \cite{Hubig2015StrictlyExpansionb} where a subspace expansion is performed on one internal leg as shown in \fir{fig:nodedirectsum}(a). Another example is performing the addition of two wave functions $|\Psi_C\rangle = |\Psi_A\rangle + |\Psi_B\rangle$ in the MPS representation. To do this each node in the first MPS network is added to the corresponding node in the second MPS network, where the dimension of both internal legs is expanded but the basis on the physical legs remains the same as illustrated in \fir{fig:nodedirectsum}(b).

\begin{figure}[ht]
\begin{center}
\includegraphics[scale=1.2]{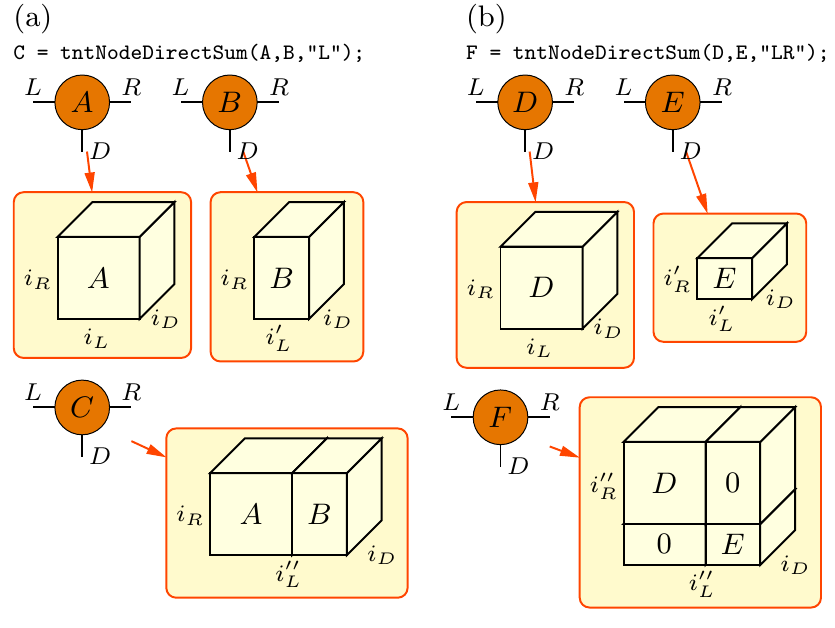}
\end{center}
\caption{Performing a direct sum of two nodes. Here the node legs $L$, $R$ and $D$ are mapped to tensor indices $i_L$, $i_R$ and $i_D$ respectively. In (a) the direct sum expands the basis on leg $L$, and the new tensor has dim(${i''}_L$) = dim(${i'}_L$)+dim(${i}_L$). The indices $i_R$ and $i_D$ should be identical. In (b) the direct sum expands the basis on legs $L$ and $R$, so the new tensor has dim(${i''}_L$) = dim(${i'}_L$)+dim(${i}_L$) and dim(${i''}_R$) = dim(${i'}_R$)+dim(${i}_R$), with zeros inserted for elements of the tensor which correspond to the initial indices ${i}_L,{i'}_R$ and ${i'}_L,{i}_R$.}
\label{fig:nodedirectsum}
\end{figure}

 \subsection{Node utility routines}
A number of utility routines are provided for performing basic operations on each \tnode. Some of the most often-used routines include:
\begin{description}
\item[\code{tntNodeCreate()}] Creates a new node, using either supplied tensor values or random tensor values. See Listing \ref{lis:qnum} for an example of creating a node using this function.
\item[\code{tntNodeCopy()}] See \fir{fig:nodecopy}. This does not copy the entire tensor but instead creates another node that points to the same data values. A conjugate copy can be taken, which simply adds a conjugate flag rather than conjugating all the values. If \tnode~operations are later applied which change the tensor values, a new deep copy of the tensor is taken.
\item[\code{tntNodePrint*()}] A set of functions that provide different formats for printing out information about a \tnode, including printing out tensor values reshaped to a matrix (with the row legs and column legs specified in the arguments) or printing out the information (e.g. leg types, dimensions, symmetry properties) only.
\item[\code{tntNodeGet*()}] These functions retrieve values from the a \tnode, for example the first value, the diagonal values, or the trace of the tensor or the size of one of the legs (i.e. dimensions).

\end{description}

\begin{figure}[ht]
\begin{center}
\includegraphics[scale=1.2]{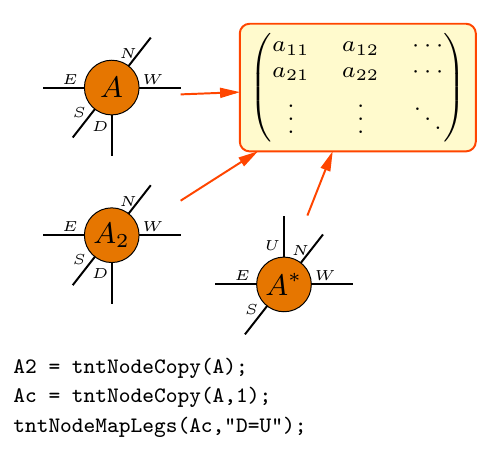}
\end{center}
\caption{Copying an original node $A$ to make an identical copy $A_2$ and a node that is the complex conjugate $A^{\ast}$. After the complex conjugate is taken the leg labels are mapped, so that $A^{\ast}$ is labelled as an upwards-facing node rather than a downwards-facing node. None of these operations change the underlying values of the tensor or the ordering of the indices, so all nodes point to the same tensor. This makes a copying a node a cheap operation.}
\label{fig:nodecopy}
\end{figure}

\subsection{Changing node connections}
There are also many functions which are not concerned with any of the tensor values, but only with the connections or properties of legs of the nodes. These include functions that allow any general network to be constructed simply by joining \tnode~legs together. In addition they include functions for modifying the properties of the \tnode~legs. Some examples of these functions are listed below.
\begin{description}

\item[\code{tntNodeJoin()}] Joins two nodes along the legs specified -- see \fir{fig:nodeop}(a) and lines 17-19 of Listing \ref{lis:varmincontract}. This means that any subsequent calls to contract which contains these nodes will result in a contraction along this index.
\item[\code{tntNodeInsert()}] Inserts a node between two nodes that are already connected. See \fir{fig:nodeop}(b).
\item[\code{tntNodeSplit()}] Removes all connections between a pair of nodes.
\item[\code{tntNodeSqueeze()}] This function removes the listed singleton legs from a node.
\item[\code{tntNodeAddLeg()}] This function adds singleton legs to a node.
\item[\code{tntNodeFuse()}] Fuses two legs together. Note this does not actually result in any change in the underlying tensor i.e.\ a reordering of indices, which could prove inefficient. See \fir{fig:nodeop}(d).
\end{description}

\begin{figure}[ht]
\begin{center}
\includegraphics[scale=1.2]{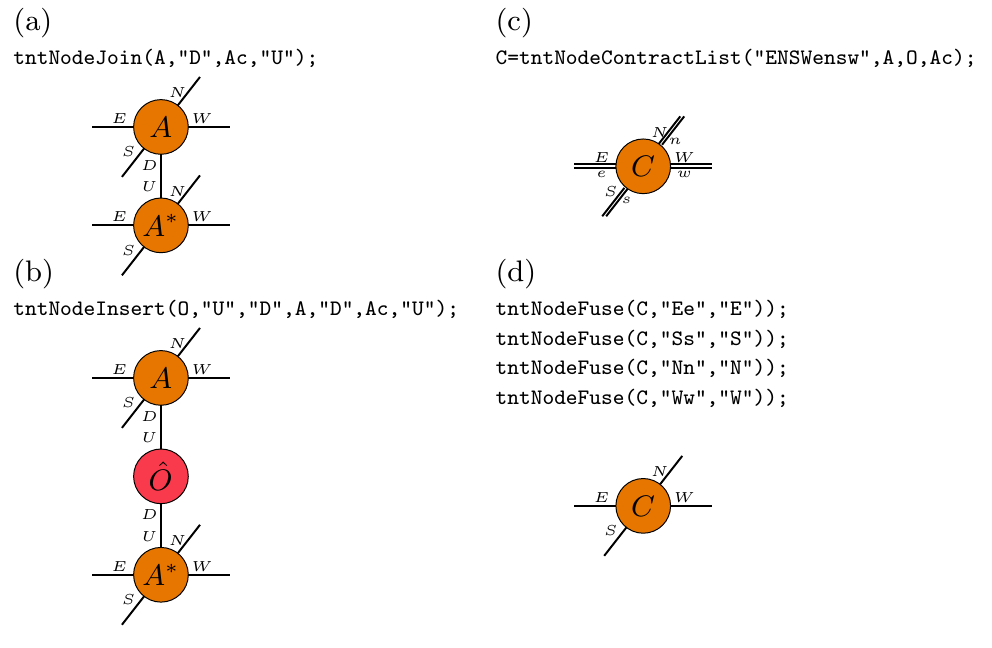}
\end{center}
\caption{Illustration of some basic operations that can be carried out on nodes. (a) Two PEPS nodes are joined along their physical legs. (b) A single-site operator node is inserted between them.
(c) The whole group of nodes are contracted leading to a node with 8 legs. (d) The legs are fused pairwise to result in a node with 4 legs.}
\label{fig:nodeop}
\end{figure}

\begin{figure*} [ht]
\begin{center}
\includegraphics[scale=1.2]{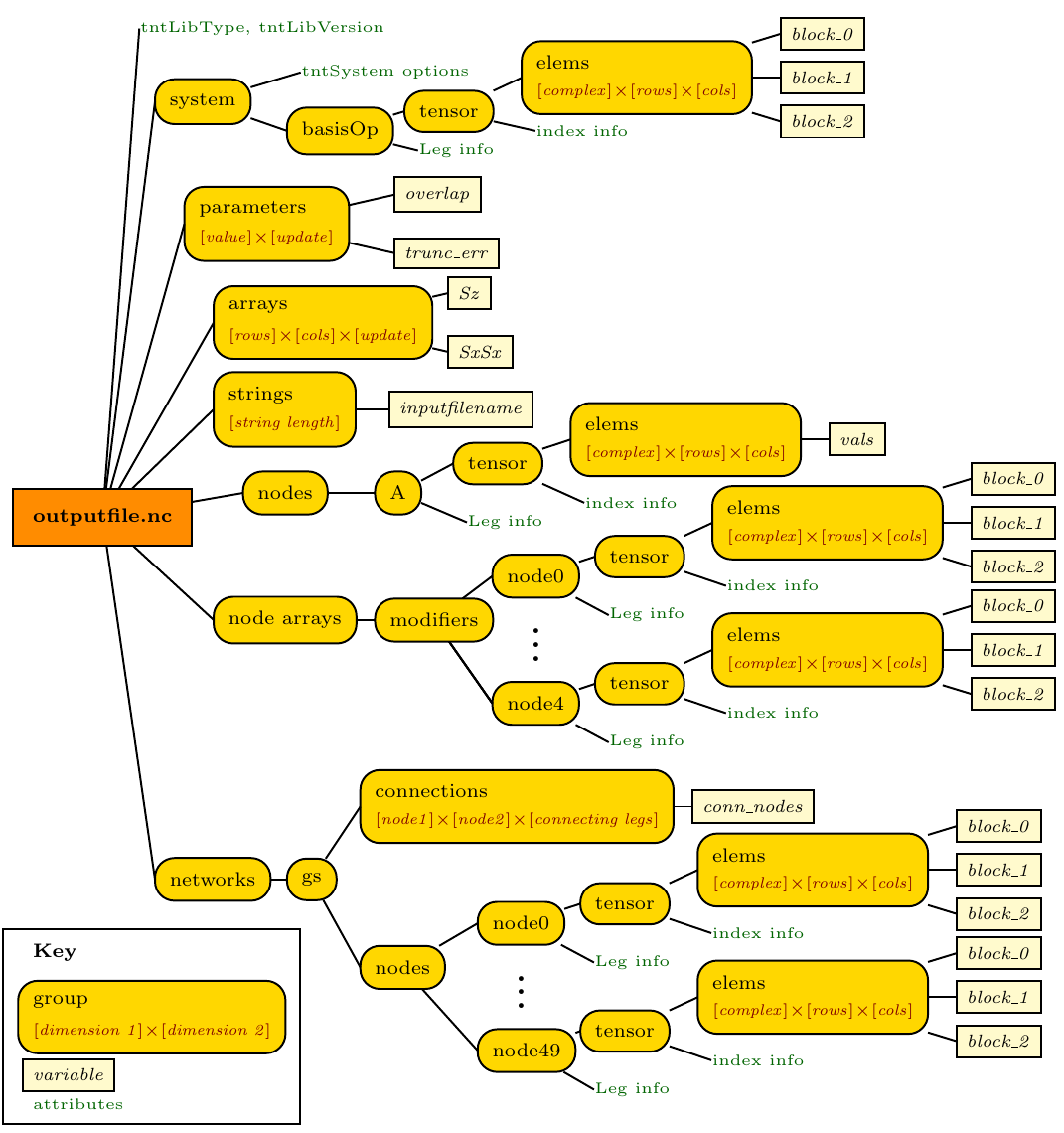}
\end{center}
\caption{The hierarchical group structure for an example output file in NetCDF format. The groups each have several attributes, and contain the variables which hold the data in the form of arrays.}
\label{fig:netcdf}
\end{figure*}

\subsection{Manipulating networks}
There are some basic functions in the Tier 1 core library for manipulating networks. This is consistent with our Tier structure since they are all functions of sufficient generality that they do not depend on the network geometry. Consequently they are limited in number, with the majority of network-level functions provided in the additional network-specific libraries.
 \begin{description}
\item[\code{tntNetworkCreate()}] Creates a new empty network. This network will contain no nodes -- nodes can subsequently be inserted using \code{tntNodeInsertAtStart()} or \code{tntNodeInsertAtEnd()}.
\item[\code{tntNetworkCopy()}] Returns a handle to a network formed of copies of all the nodes in the original network, and all copies connected in the same way as in the original network. Like \code{tntNodeCopy()} these copies do not copy all the tensor values, but instead create additional pointers to the tensors. It is also possible to create a copy of a network with the complex conjugate of all nodes taken.
\item[\code{tntNetworkSplit()}] Splits a network into two separate networks. All the nodes involved in the split must be given.
\item[\code{tntNetworkToNodeGroup()}] Deletes the network information (i.e.\ the network structure, and any network information) but leaves all connections between \tnode s that formed the network intact. This can be useful when joining two networks together to create a single network. It can also be useful when the entire network is contracted to a single node (for an example see line 14 of Listing \ref{lis:varmin}) such that a network handle is no longer necessary.
\end{description}

\subsection{System settings} \label{sec:systemfunctions}
There are a number of functions that can be used to change global calculation parameters, for example those used when performing a truncated SVD. For the full list please see the documentation \cite{TNTdocs} -- only those related to functions already described above are given here.
 \begin{description}
\item[\code{tntSysInfoPrint()}] Prints all the current system parameters to the standard outputs. See Listing \ref{lis:sys} for an example of the output.
\item[\code{tntSVDTruncTolSet()}] During an SVD, all singular values $\lambda_i$ less that this will be discarded.
\item[\code{tntSVDRelTruncTolSet()}]During an SVD, all scaled singular values $\lambda_i/\lambda_0$ less that this will be discarded.
\item[\code{tntSVDTruncErrTolSet()}]During an SVD, the maximum number of singular values will be discarded for which the truncation error is still less than this bound.
\item[\code{tntSVDTruncType()}]Set the function used to calculate the truncation error for all SVDs.
\item[\code{tntSVDTolSet()}]Set the tolerance for zeroing values during the SVD (see \secr{sec:performance}).
\end{description}

\subsection{Input and output} \label{sec:io}
The library supports input and output of data in MATLAB and NetCDF format, and MATLAB scripts are provided to convert between the two formats. In both cases the input and output data is richly structured to reflect the complex data that can exist in the library. As described above these structures are necessary to keep track of labelling and ordering of legs and indices in the simplest case, or more complicated block structures for symmetric nodes, as well as connections in networks of any general geometry. In addition all output files will contain the \tsys~structure and information about the library version. When such an output file is later used as an input file, the \tsys~structure is automatically loaded, ensuring that the simulation proceeds with the same system parameters (e.g.\ truncation tolerances, symmetry information).

MATLAB provides a convenient format and is already widely used within the community. However MATLAB is not suitable for all applications and/or users since it is a commercial package. Furthermore some computing resources, such as the UK National supercomputer ARCHER, do not have MATLAB installed as standard.

NetCDF 4 has been chosen as an alternative data format since it is also widely used within the scientific community, is freely available \cite{Unidata4}, and is already installed on many computing facilities. Like HDF5 (another commonly used scientific data format) it supports hierarchical group structures and is self-describing. However the interface for NetCDF 4 is considerably simpler whilst still containing all the flexibility required for description of TNT data structures (see \fir{fig:netcdf}).

For MATLAB output, nodes and networks are represented by means of custom `structure'-type variables which are created by functions in a separate MATLAB TNT library. Example initialisation scripts using these functions are available with the Tier 3 applications described in \secr{sec:tier3}.

\section*{Acknowledgements}
The authors would like to acknowledge the use of the University of Oxford Advanced Research Computing (ARC) facility \cite{richardsarc} in carrying out this work. S.A. and D.J. acknowledges support from the EPSRC Tensor Network Theory grant (EP/K038311/1).\\

\bibliographystyle{unsrt}
\bibliography{tntpaper}

\end{document}